# Controlling Surface Chemistry in Tin-Lead Perovskites for High-Efficiency Solar Cells

Lijun Chen[1], Sitong Zhao[1], Gaukhar Nigmetova[1], David Garcia Romero[1], Matteo Pitaro[1], Sofia Lerda[2,3], Edoardo Mosconi[3], Michael Aidi[1], Douwe van Netten[1], Julia Kraft[1], Jiaxiong Li[1], Jacopo Pinna[1], Asya Serafimova[1], Antonija Grubišić-Čabo[1], Loredana Protesescu[1], Giuseppe Portale[1], Petra Rudolf[1], Filippo De Angelis[2,3,4], Maria Antonietta Loi[1]*

1. Zernike Institute for Advanced Materials, University of Groningen, Nijenborgh 3, 9747 AG Groningen, The Netherlands
2. Department of Chemistry, Biology and Biotechnology, University of Perugia, INSTM, Via Elce di Sotto 8, Perugia 06123, Italy
3. Computational Laboratory for Hybrid/Organic Photovoltaics (CLHYO), Istituto CNR di Scienze e Tecnologie Chimiche "Giulio Natta" (CNR-SCITEC), Via Elce di Sotto 8, Perugia 06123, Italy
4. SKKU Institute of Energy Science and Technology (SIEST), Sungkyunkwan University (SKKU), Suwon 16419, Republic of Korea
*Corresponding author. Email: m.a.loi@rug.nl

## Abstract
Interfacial nonradiative recombination limits tin-lead perovskite photovoltaics. Here we show that neutral ethylenediamine chelates undercoordinated metals at tin-rich surfaces, weakens metal-iodide bonds, and removes excess tin, reconstructing the surface towards a stoichiometric tin-to-lead ratio. In contrast, protonated ethylenediammonium iodide initially passivates undercoordinated metal sites before undergoing defect-assisted deprotonation. Atomistic simulations reveal that subsurface iodide vacancies promote iodide migration and lower the energetic cost of deprotonation. Ethylenediamine-induced reconstruction produces stronger n-type surface character, facilitates electron extraction into $C_{60}$, and suppresses interfacial nonradiative recombination, enabling methylammonium-free solar cells with a power conversion efficiency of 25.86%. Together, these distinct surface reaction pathways and the defect-assisted evolution of the passivating species underscore the importance of understanding perovskite surface chemistry for the rational design of molecular passivation strategies.

## Introduction
All-perovskite tandem solar cells have significantly surpassed the power conversion efficiency (PCE) records of single-junction devices.[1,2] High-performance narrow-bandgap tin-lead (Sn-Pb) perovskite solar cells (PSCs) are critical for the realization of efficient all-perovskite tandem devices.[3,4,5,6,7] Recent achievements in PCE exceeding 24% have been driven primarily by suppressing perovskite surface traps and reducing recombination losses at the perovskite/charge transport layer interface.[2,8,9] In particular, recombination losses at the buried perovskite/hole transport layer interface have been largely mitigated through the development of hole-selective self-assembled monolayers[10,11,12,13] and related interfacial modification strategies.[2,14] However, despite extensive efforts, the top interface between the perovskite absorber and the electron transport layer (fullerene $C_{60}$) remains a critical source of nonradiative recombination.

Recombination losses mainly originate from exposed perovskite surfaces with a high density of trap states, which hinder charge transport and extraction and result in a loss of open-circuit voltage and fill factor. Consequently, Sn-Pb perovskite solar cells still underperform their pure-Pb counterparts. To address these issues, various strategies have been proposed, including functionalized fullerene derivatives, Lewis-base molecules and organic ammonium salts.[15] Among these approaches, diammonium and diamine molecules have been explored for the interfacial passivation of Pb[16,17,18] and Sn-Pb based perovskite surfaces.[19,20,21,22,23] However, the chemical identity of the active passivating species remains elusive, and the molecular mechanisms governing their interactions with the perovskite surface and their effects on nonradiative recombination and interfacial electronic properties are still not fully understood. This uncertainty is reflected in the varying outcomes reported for nominally similar passivation treatments across different perovskite compositions and processing conditions. Even the closely related ethylenediamine (EDA) and ethylenediammonium diiodide ($EDAI_2$) treatments have been reported to produce different effects.[17,23,24,25,26] Establishing how molecular protonation governs surface chemistry is therefore critical to developing rational passivation strategies for high-performance Sn-Pb perovskite photovoltaics.

Here, we combine a (2-(3,6-Dibromo-9H-carbazol-9-yl)ethyl)phosphonic acid (Br-2PACz)-based buried interface with EDA treatment of the top interface in methylammonium (MA)-free Sn-Pb perovskite solar cells. Direct comparison of EDA and $EDAI_2$, supported by surface-sensitive measurements and atomistic simulations, reveals distinct surface reaction pathways. Mechanistically, neutral EDA chelates excess surface Sn species and drives reconstruction towards a near-stoichiometric Sn-to-Pb ratio, thereby homogenizing the surface energy landscape, whereas $EDAI_2$ undergoes defect-assisted deprotonation and induces substantially less Sn removal and surface reconstruction. EDA-treated devices achieve a champion PCE of 25.86%, compared with 18.11% for untreated references and 23.86% for $EDAI_2$-treated devices, while retaining more than 90% of their initial PCE after 600 h at 65 °C. Therefore, these results establish record performance in MA-free Sn-Pb perovskite solar cells while providing a microscopic understanding of the surface chemistry governing device efficiency.

**Influence of passivation molecules on perovskite surface morphology and crystallinity**

Sn-Pb perovskite ($FA_{0.75}Cs_{0.25}Pb_{0.5}Sn_{0.5}I_3$) films were deposited on Br-2PACz-functionalized indium tin oxide substrates, as previously reported by our group.[11] The un-passivated perovskite reference film (Fig. 1a) exhibited well-defined crystalline grains with sharp boundaries, along with several bright crystalline features which are assigned to $PbI_2$/$SnF_2$/$SnI_2$ crystallites located at the perovskite grain boundaries.[19] After surface modification with $EDAI_2$, most of these bright features disappeared, probably due to the combined effect of the passivating agent and the solvent isopropyl alcohol (IPA). Notably, IPA alone influenced the surface morphology[27] but did not fully remove the bright crystalline features (Fig. S1). In contrast, EDA treatment induced pronounced restructuring of the grains, with the grain boundaries appearing partially dissolved and the surface evolving into a terraced morphology (Fig. 1c). We have previously reported that amine-containing molecules can selectively cleave specific crystal planes {100}, leading to strong surface restructuring in $MAPbBr_3$ single crystals.[28] Atomic force microscopy (AFM) images (Fig. S2) show a slight increase of root-mean-square (RMS) surface roughness from 24.1 nm for the reference film to 26.2 nm after $EDAI_2$ treatment. EDA treatment reduced the roughness even further to 20.3 nm, and yielded a morphology characterized by grain merging and large-scale surface flattening. In addition, small surface protrusions are observed, indicating local surface restructuring (Figs. S3 and S4). Correlation between the height and peak-force error channel in atomic force microscopy further confirms that the EDA-induced features are not associated with grain boundaries but instead arise from a large-area surface restructuring, distinctly different from the effect induced by $EDAI_2$ (Figs. S2b and S5).

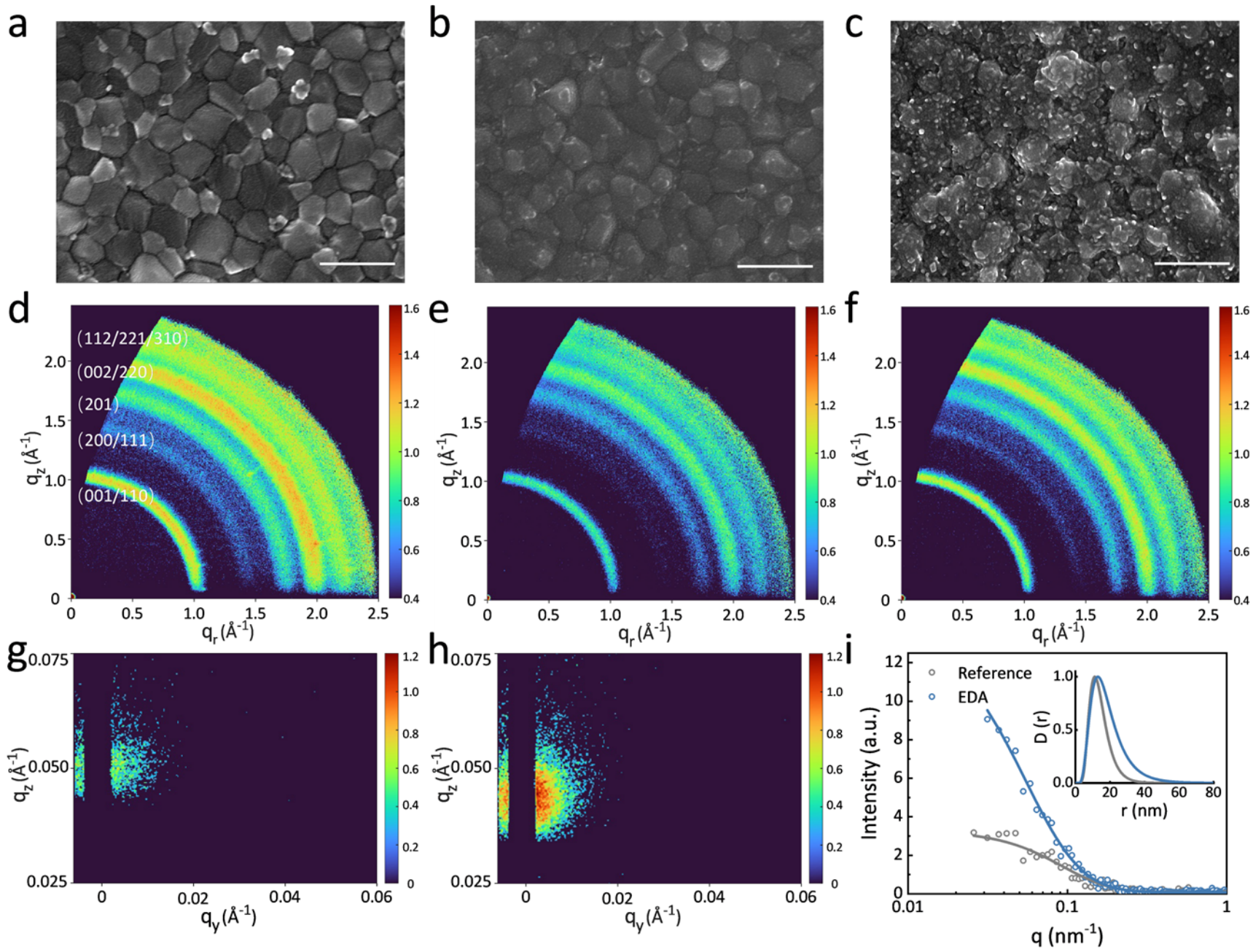

**Fig. 1 Surface morphology and crystallinity of Sn-Pb perovskite films.** SEM images of un-passivated reference (a), $EDAI_2$-passivated (b), and EDA-passivated (c) Sn-Pb perovskite films. Scale bar: 1 µm. (d-f) GIWAXS patterns for the corresponding perovskite samples acquired with an incident angle of 0.4° (surface region). Zoomed view of the GISAXS patterns for the (g) reference and (h) EDA-treated films. (i) GISAXS horizontal intensity cuts for the reference and EDA-treated films. Solid lines denote the best fits obtained using a polydisperse spherical particle model. Inset: corresponding particle size distributions extracted from the fits.

To gain deeper insight into the passivation-induced crystallinity changes on bulk and surface of the perovskite, we conducted grazing-incidence wide-angle X-ray scattering (GIWAXS) and X-ray diffraction (XRD) measurements. All films retained the same pseudo-cubic tetragonal structure, in agreement with previous reports,[29] with no discernible changes in the bulk XRD patterns following EDA or $EDAI_2$ treatment (Figs. S6-S8; Tables S1 and S2; Supplementary Text 1). By contrast, surface-sensitive GIWAXS (0.4°) measurements revealed reduced ring broadening, slight preferential texturing, and variations in peak position as well as relative peak intensities (Fig. 1d-f), whereas these differences became negligible under bulk-sensitive conditions (4°; Fig. S9), confirming that the structural modifications are largely confined to the near-surface region.

The reduced peak broadening in GIWAXS, with the full width at half maximum (FWHM) decreasing by ~0.02 $Å^{-1}$ (EDA) and ~0.01 $Å^{-1}$ ($EDAI_2$) (Fig. S10) after treatment, indicates enhanced structural order and more extended crystallinity, likely due to lateral grain coalescence following surface passivation, in agreement with SEM observations. Variations in relative peak intensities may arise from differences in texturing, site occupancy, or atomic positions. Quantitative azimuthal analysis of the (001)/(110) and (201) diffraction rings (Fig. S11) reveals negligible texturing in the reference film, whereas EDA treatment induced weak preferential orientation, evidenced by enhanced (201) intensity at 40-50° and increased (001/110) intensity at 0°, suggesting alignment of (001/110) planes parallel to the substrate. In contrast, $EDAI_2$-treated films exhibited isotropic distributions, indicating passivation-dependent structural changes. Depth-resolved GIWAXS (Fig. S12) shows pronounced discrepancies between surface and bulk in the un-passivated reference film, including peak broadening, shifts to lower q-values, and asymmetric features (Fig. S12a), which are substantially suppressed after EDA treatment (Fig. S12b). This indicates that surface treatment promotes vertical structural homogeneity by eliminating off-stoichiometric species.

In agreement with SEM observations, the GISAXS pattern of the EDA-treated sample showed an enhanced signal, consistent with a larger number of nanometer-sized scatterers than in the reference sample (Figs. 1g and 1h). Information about the lateral particle size was extracted from the fit of the intensity cuts along the horizontal $q_y$ direction for the samples (Fig. 1i). The average dimensions of the small protrusions obtained from the best fit of the GISAXS data are ~20 nm and ~26 nm for reference and EDA-treated films, respectively, consistent with the size observed by AFM (Fig. S3b).

**Passivation-induced changes in Sn-Pb perovskite surface chemistry**

We next used X-ray photoelectron spectroscopy (XPS) to determine whether the passivation merely heals pre-existing defects or physically reconstructs the surface and consequently also alters the chemical composition, as suggested by the SEM images (Fig. 1c). The detailed core-level spectra are shown and explained in the Fig. 2a-c and Supporting Information (Figs. S13 and S14; Tables S3 and S4; Supplementary Text 2). The most pronounced differences among the reference, $EDAI_2$-treated and EDA-treated surfaces are observed in the N 1*s* core-level spectra (Fig. 2a). The un-passivated film exhibits a single component at a binding energy (BE) of 400.6 eV, assigned to $FA^+$. Following EDA treatment, an additional component appears at 399.9 eV, corresponding to the $R\text{-}NH_2$ groups of EDA. The $EDAI_2$-treated surface similarly exhibits components assigned to $FA^+$ and $R\text{-}NH_2$ at 400.4 and 399.7 eV, respectively. Notably, neither treated surface shows a detectable signal at approximately 402.1 eV, where protonated amine groups (R-$NH_3^+$) would be expected. These results demonstrate that neutral amine species are present on both treated surfaces, even when $EDAI_2$ is used as the precursor, indicating substantial surface-induced deprotonation of the diammonium species.

Quantitative XPS analysis further reveals pronounced differences in the surface compositions (Supplementary Text 2 and Tables S3 and S4). The un-passivated surface is highly non-stoichiometric, with $Cs^+$/formamidinium ($FA^+$) and $Sn^{2+}/Pb^{2+}$ ratios of 0.7 and 2.5, respectively, probably owing to surface phase segregation[30] and imbalanced crystallization kinetics.[31,32,33,34] Iodide is also distributed unevenly between the two metal species: $I^-/Sn^{2+}$ and $I^-/Pb^{2+}$ ratios are 3.0 and 7.5, respectively, while the overall $I^-/(Sn^{2+} + Pb^{2+})$ ratio is only 2.2. These values indicate that the untreated surface is enriched in Cs and Sn and contains a substantial population of iodide-deficient, undercoordinated $Sn^{2+}$ species.

EDA treatment markedly reconstructs the surface composition (Supplementary Text 2). The $Sn^{2+}/Pb^{2+}$ ratio decreased from 2.5 to 1.2 (Fig. 2d), approaching the nominal value of 1.0, while the $I^-/(Sn^{2+} + Pb^{2+})$ ratio increased from 2.2 to 3.2. The $Cs^+/FA^+$ ratio increased from 0.7 to 0.9 (Tables S3 and S4), indicating a relative depletion of FA-containing species, consistent with the GIWAXS peak shift to higher q-values (Fig. S12).[29] The preferential interaction of EDA with $Sn^{2+}$ is consistent with the hard-soft acid-base (HSAB) principle: $Sn^{2+}$, a hard Lewis acid, preferentially binds to amine ligands via hard-hard interactions, while $Pb^{2+}$, a softer acid, favors iodine, as shown by higher-order iodine species in Pb-based complexation.[35] Liquid-state $^{119}Sn$ nuclear magnetic resonance (NMR) of dissolved pristine films in deuterated solvent revealed a deshielded $^{119}Sn$ signal upon EDA addition (Figs. S15 and S16), supporting Sn-EDA complexation. Although the measurements do not distinguish between bulk and surface species, together with the XPS results they support a mechanism in which EDA preferentially coordinates excess surface Sn and facilitates its removal, accompanied by redistribution of FA-containing species and reconstruction towards a near-stoichiometric surface.

$EDAI_2$ produces a markedly different compositional response (Tables S3 and S4). The surface $Sn^{2+}/Pb^{2+}$ ratio decreased slightly from 2.5 to 2.1 (Fig. 2d), and the $I^-/(Sn^{2+} + Pb^{2+})$ ratio increased to 2.6, remaining below that obtained with EDA. These comparatively small changes indicate substantially less Sn removal and more limited surface reconstruction. By contrast, the $Cs^+/FA^+$ ratio increased sharply from 0.7 to 1.8, revealing pronounced depletion or redistribution of FA-containing surface species. Because $Cs^+$ is comparatively non-volatile, this change is most likely determined by the FA extraction promoted by iodide-assisted acid-base chemistry during $EDAI_2$ deprotonation. Together with the absence of an $R\text{-}NH_3^+$ signal in the N 1*s* spectrum, these observations suggest that $EDAI_2$ undergoes deprotonation during surface treatment and annealing (100 °C). The Lewis acidity of surface $Sn^{2+}$ species, together with the slight shift of all core-level peaks toward lower binding energies, suggests an increase in electron density that facilitates proton abstraction and hydroiodic acid (HI) formation. The resulting neutral amine species can subsequently coordinate with excess surface $Sn^{2+}$, but the limited decrease in $Sn^{2+}/Pb^{2+}$ indicates substantially less Sn removal than with direct EDA treatment. We propose that sequential deprotonation and complexation limit Sn removal, resulting in less extensive surface reconstruction than with direct EDA treatment.

Furthermore, consumption and volatilization of iodide-containing species during HI formation likely limit the increase in surface iodide concentration. Given the XPS probing depth of approximately 6-8 nm (Table S5 and Supplementary Text 3), these compositional changes predominantly reflect restructuring of the near-surface region. Together, these results point to distinct interfacial reaction pathways for neutral EDA and protonated $EDAI_2$. The defect-assisted origin and energetics of the deprotonation pathway are examined below using atomistic simulations.

In addition, Fourier-transform infrared spectroscopy (FTIR) and XRD measurements revealed no evidence of imidazolium formation following EDA treatment of Sn-Pb perovskites, suggesting a distinct surface chemistry from that reported for Pb-based counterparts (Figs. S17 and S18; Supplementary Text 4).[17] Moreover, EDA stabilized $Pb_{0.5}Sn_{0.5}I_2$, whereas $EDAI_2$ induced pronounced chemical instability, further supporting stable EDA interactions with the Sn-Pb perovskite (Figs. S19 and S20).

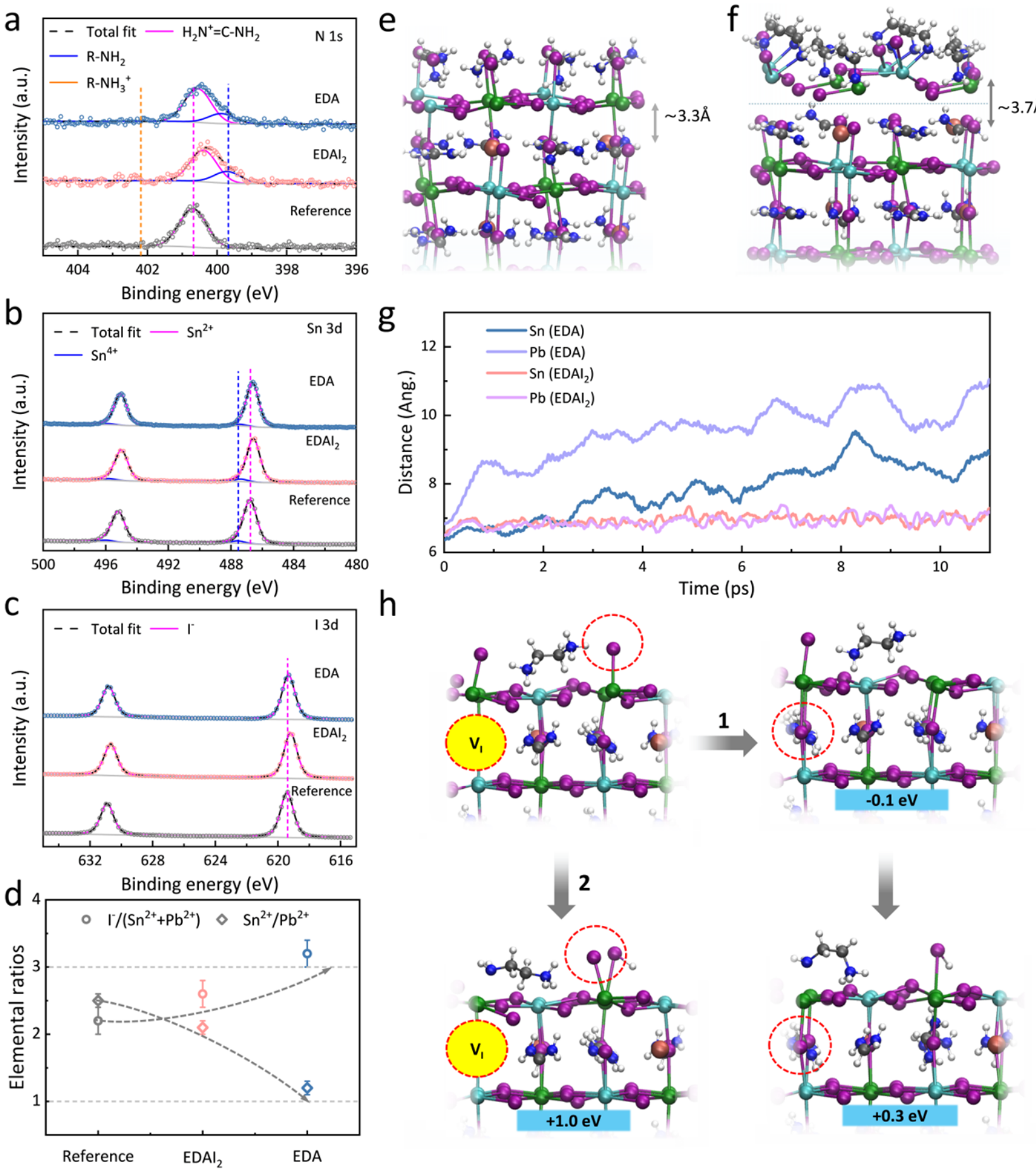

**Fig. 2 XPS and computational investigation of Sn-Pb perovskite surface chemistry.** XPS spectra of the (a) N 1*s*, (b) Sn 3*d*, and (c) I 3*d* core-level regions (dots) of the reference, $EDAI_2$-passivated, and EDA-passivated Sn-Pb perovskite films and corresponding fits (solid lines). (d) Elemental ratios derived from the atomic percentages of the various Sn-Pb perovskite films. Geometry structure of the (e) $EDAI_2$ and (f) EDA passivated slabs. (g) Evolution of the average distance between the surface metal atoms (Sn and Pb) and the underlying lattice layer along the molecular dynamic simulation time for the EDA and $EDAI_2$ passivated model, respectively. (h) Molecular simulation of the deprotonation process of $EDAI_2$ on top of the Sn-Pb in presence of iodine vacancy ($V_I$).

Following the experimental observations, we investigated the atomistic mechanisms of EDA and $EDAI_2$ on the metal-terminated mixed Sn-Pb surface.

Starting from the pristine surface, we modelled the interaction of a single EDA or $EDAI_2$ species with undercoordinated metal sites to identify their preferred adsorption configurations. For EDA, the adsorption energy was calculated in a chelating geometry through the lone pairs of its two amino groups, whereas $EDAI_2$ was investigated in a conventional ammonium-based configuration, resembling the organic-inorganic arrangement generally observed at passivated perovskite surfaces (Supplementary Text 5). Both species preferentially interact with Sn over Pb, with an energetic preference of approximately 0.2 eV for EDA and 0.1 eV for $EDAI_2$, indicating stronger interactions with undercoordinated Sn sites.

We subsequently investigated the limiting case of full surface coverage (Figs. 2e, f), EDA forms a dense chelating layer that coordinates exposed surface metal sites, whereas $EDAI_2$ adopts an ammonium-based configuration, with protonated amino groups and iodide counterions interacting with the inorganic surface and saturating the undercoordinated metal sites. Electronic-structure analysis of the fully passivated surfaces showed that both treatments effectively suppress the trap states associated with undercoordinated Pb and Sn atoms, restoring a clean band-edge electronic structure (Fig. S21a-c). However, their passivation mechanisms differ substantially at the atomic level. $EDAI_2$ forms a stable passivating overlayer that electronically neutralizes the undercoordinated surface metal sites (Fig. 2e). In contrast, EDA chelates exposed Pb and Sn atoms, inducing substantial restructuring of the inorganic surface while removing the associated electronic trap states (Fig. 2f). Thus, the coordination-driven mechanism is fundamentally different from the predominantly electrostatic stabilization provided by the diammonium species.

Notably, EDA-chelation substantially elongated the metal-iodide bonds connecting surface Pb and Sn atoms to the underlying inorganic lattice (Figs. 2e and 2f). The resulting weakening of the anchoring bonds suggests that EDA coordination may facilitate the detachment and eventual removal of undercoordinated surface metal species. To further investigate this behavior under finite-temperature conditions, we performed Car-Parrinello molecular dynamics simulations on the full-coverage surfaces passivated with $EDAI_2$ and EDA. The evolution of the distance between the surface metal atoms and the underlying inorganic layer is reported in Fig. 2g. The $EDAI_2$-passivated interface remained structurally stable throughout the simulated time window, and the diammonium species preserved their interaction with the perovskite surface without inducing significant disruption of the underlying inorganic framework. However, upon thermal activation, the EDA chelated surface metal atoms progressively detached from the perovskite lattice (Fig 2g). This behavior is consistent with the pronounced elongation and weakening of the metal-iodide bonds observed in the optimized structures (Figs. 2e and 2f).

However, although the simulations predicted a stable $EDAI_2$ overlayer, XPS revealed a clear signal associated with neutral amine groups but no detectable ammonium species. We therefore investigated the atomistic mechanism and energetics of $EDAI_2$ deprotonation at the perovskite surface. Starting from the adsorption geometry of $EDAI_2$ on the metal-terminated surface, we considered a configuration in which the two iodide counterions coordinate undercoordinated surface Sn atoms, while the protonated amino groups are stabilized by a hydrogen-bonding network with surface halides. We examined deprotonation both on the pristine surface and in the presence of a subsurface iodine vacancy located beneath the Sn atom interacting with the adsorbed passivating species (Fig. 2h). On the defect-free surface, deprotonation is energetically unfavorable, with a reaction energy of approximately +0.8 eV. And the energy penalty increases to approximately +1.0 eV (path 2) when a subsurface iodine vacancy is present but remains un-passivated, demonstrating that the vacancy alone is insufficient to promote deprotonation.

A markedly different behavior emerges when migration of one of the surface iodide ions is allowed (Fig. 2h). An iodide initially associated with the adsorbed $EDAI_2$ migrates into the first perovskite layer and fills the underlying iodine vacancy, stabilizing the defective lattice by ~0.1 eV and substantially modifying the energetics of the subsequent proton-transfer reaction. Following vacancy healing, the deprotonation energy decreases to ~ +0.3 eV (path 1), reducing the energetic penalty by ~0.5 eV relative to the pristine surface, demonstrating that iodine vacancies can substantially facilitate the conversion of diammonium species into neutral EDA (Fig. 2h).

These results reveal a defect-assisted mechanism in which subsurface iodine vacancies not only represent electronically detrimental defects requiring passivation but also actively mediate surface chemistry by promoting iodide migration and lowering the energetic cost of $EDAI_2$ deprotonation. This mechanism provides a consistent atomistic explanation for the neutral diamine species observed by XPS, even when $EDAI_2$ is employed as the molecular precursor. The resulting EDA can subsequently chelate undercoordinated surface metals, linking vacancy healing and molecular conversion to electronic passivation and the removal of weakly bound Pb and Sn species.

**Sn-Pb perovskite surface recombination losses and energy band offset**

Kelvin probe force microscopy (KPFM) mapping revealed a treatment-induced homogenization of the surface-potential landscape (Figs. 3a-c) of the Sn-Pb perovskite film. The standard deviation of the contact potential difference ($V_{CPD}$) decreased from 12.1 mV for the reference to 10.7 and 11.1 mV for the $EDAI_2$- and EDA-treated films (Figs. 3d-f), respectively, indicating reduced potential fluctuations and a more homogeneous interfacial contact with $C_{60}$.[36] At the same time, the average $V_{CPD}$ shifted progressively from -0.102 V for reference to −0.198 V and −0.263 V after $EDAI_2$ and EDA treatment, respectively, evidencing a substantial modulation of surface electrostatics towards a more n-type-like surface character,[37,38] consistent with surface compositional and structural restructuring described above. This simultaneous homogenization and electrostatic shift suggest that diamine treatment redefines interfacial energetics, facilitating efficient electron extraction into $C_{60}$ and suppressing interfacial recombination.[39] In Figs. 3g and S22, impedance spectra show a pronounced shift of the characteristic peak towards lower frequencies and an increase in peak magnitude, showing a substantial enhancement of the recombination resistance and a prolonged carrier lifetime (from 2.8 µs for the reference to 6.4 µs for the $EDAI_2$-treated and 9.5 µs for the EDA-treated perovskite film), highlighting the effectiveness of the EDA treatment in suppressing non-radiative recombination.

We further investigated the effect of passivation induced surface restructuring on carrier lifetime of Sn-Pb perovskites using photoluminescence (PL) spectroscopy. As shown in Fig. 3i, the steady state PL spectrum of the reference film exhibits a peak at ~990 nm, which slightly blue shifts after $EDAI_2$ treatment and more markedly after EDA treatment, consistent with changes in the near-surface electronic environment induced by surface restructuring. Although a blue shift in Sn-based perovskites is generally associated with doping,[40] the EDA-treated film exhibits minimal $Sn^{4+}$ formation (see Supplementary Text 2), together with a substantially prolonged carrier lifetime (Fig. 3h), suggesting that doping is unlikely to dominate the observed spectral shift. Time-resolved photoluminescence (TRPL) measurements (Fig. 3h) reveal an average lifetime of 1.64 µs for EDA-treated film, markedly longer than those of the $EDAI_2$-modified (0.65 µs) and un-passivated reference (0.61 µs) films. This lifetime enhancement indicates suppressed non-radiative recombination, consistent with effective trap-state passivation driven by the EDA-induced surface restructuring.[41] These results collectively demonstrate that diamine-based passivation is substantially more effective than its diammonium counterpart.

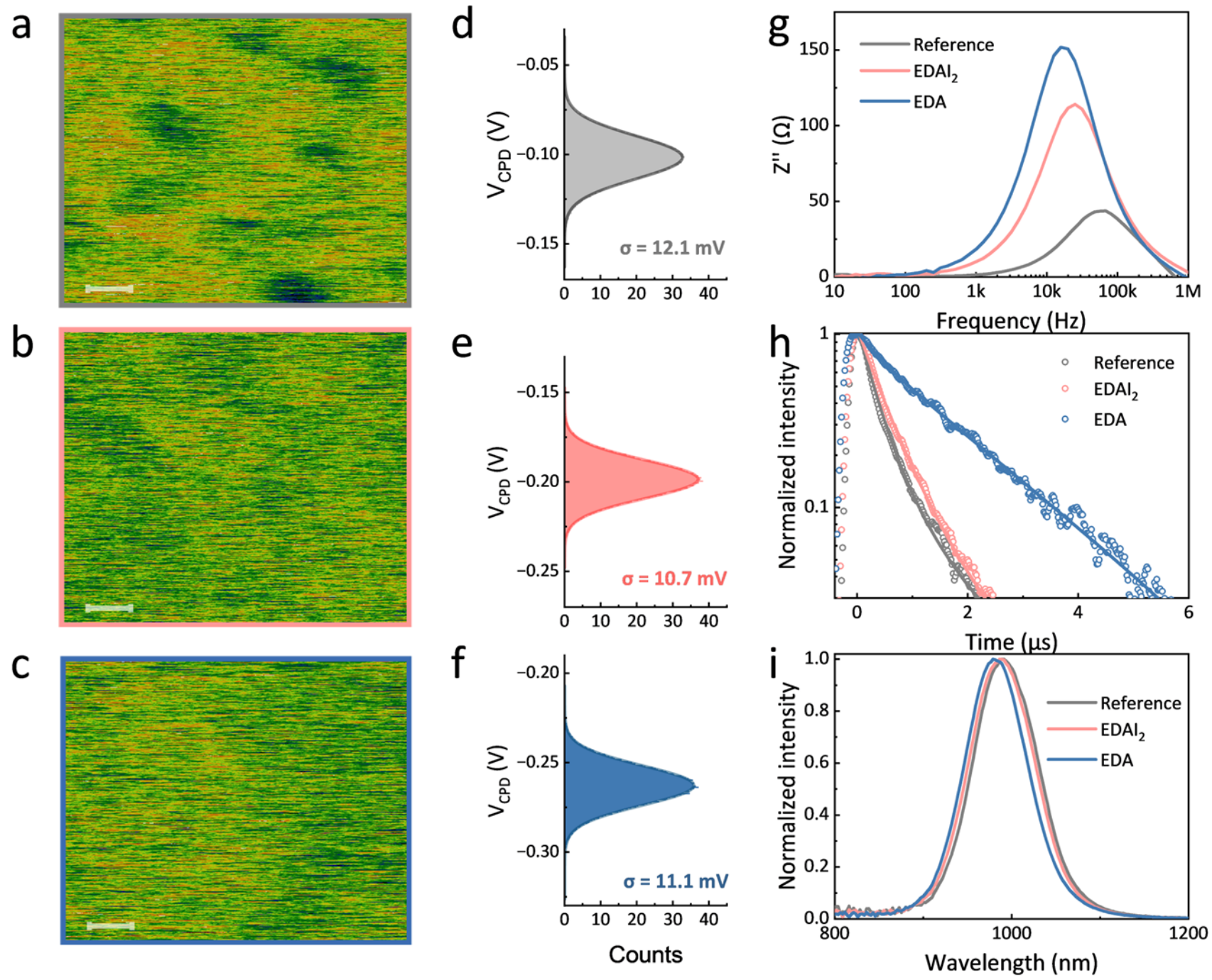

**Fig. 3 Surface-passivation-induced modulation of interfacial energetics and non-radiative recombination in Sn-Pb perovskites.** KPFM maps of reference (a), $EDAI_2$-treated (b), and EDA-treated (c) samples. Corresponding spatial fluctuations in $V_{CPD}$ (d, e, f). (g) Frequency dependence of the imaginary impedance (Z'') for the corresponding perovskite devices. TRPL (h) and PL (i) spectra of reference, $EDAI_2$-treated, and EDA-treated Sn-Pb films deposited on glass.

**Sn-Pb perovskite solar cell performance**

Fig. 4a shows the cross-sectional scanning electron microscopy (SEM) image of the Sn-Pb PSCs device, consisting of indium tin oxide (ITO)/Br-2PACz/$FA_{0.75}Cs_{0.25}Pb_{0.5}Sn_{0.5}I_3$/passivation layer/$C_{60}$/bathocuproine (BCP)/Ag, with a perovskite thickness of approximately 700 nm. The J-V characteristics of the devices under 1 sun illumination are shown in Fig. 4b. The device with the $EDAI_2$-treated perovskite surface achieved a higher PCE of 23.86% than the un-passivated reference film that reached 18.11%. The EDA passivation further enhanced photovoltaic performance, increasing the open-circuit voltage ($V_{OC}$) from 0.716 V to 0.901 V, and the fill factor (FF) from 73% to 81%, resulting in a record PCE of 25.86% which surpasses current state-of-the-art Sn-Pb PSCs (Fig. 4e and Table S6). It is important to underline that the EDA-passivated device shows well-behaved dark J-V characteristics (Fig. S23) and negligible hysteresis under illumination (Fig. S24 and Table S7). The statistical distribution of device parameters from about 40 devices (Fig. 4d and Table S8) further confirms the reproducible performance enhancement. The external quantum efficiency (EQE) spectra of the champion device yielded an integrated current of 33.28 mA $cm^{-2}$ (Fig. 4c), matching the J-V-derived current when accounting for a spectral mismatch factor of 1.07 (see supplementary Text 6 and Fig. S25). Such remarkable improvements in all the device characteristics are primarily attributed to the surface restructuring (see above discussion), with consequent reduction of non-radiative recombination losses at the perovskite/$C_{60}$ interface. EDA passivation not only improved the open-circuit voltage, which is an indication of reduced trap induced recombination, but also enhanced the fill factor (FF) to over 80%. The latter is attributed to an optimized internal electric-field distribution resulting from the surface stoichiometry re-equilibration induced by EDA passivation.[42]

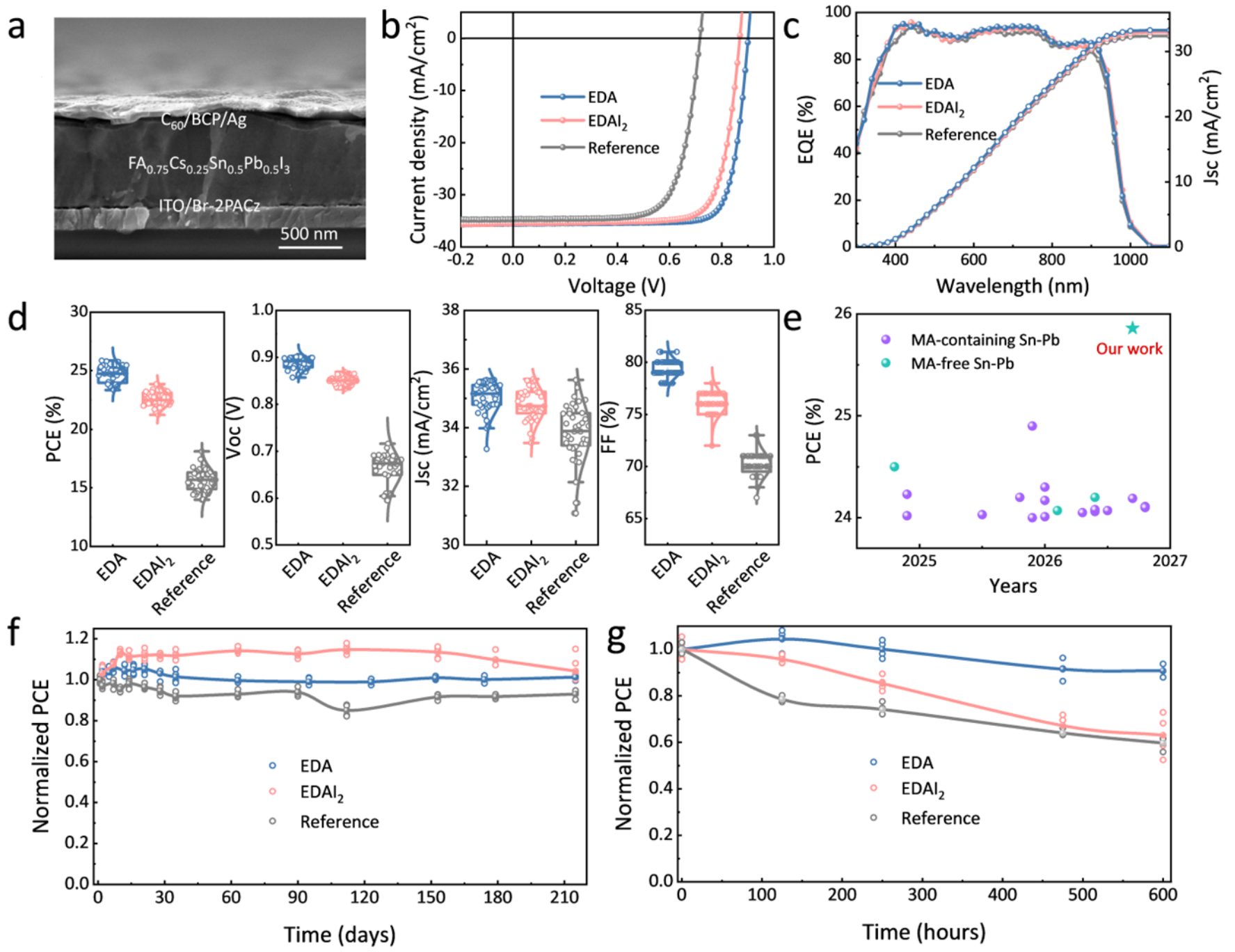

**Fig. 4 Performance of Sn-Pb devices with different surface passivation.** (a) Cross-sectional SEM image of the device. (b) J-V characteristics of the champion device measured under AM 1.5G illumination. (c) EQE spectrum of the corresponding device, with the integrated current density. (d) Statistical distribution of photovoltaic parameters, including PCE, FF, $V_{OC}$, and $J_{SC}$, based on about 40 devices. (e) Summary of reported PCEs (>24%) in literature for MA-free and MA-containing Sn-Pb perovskite solar cells, highlighting the ~1 percentage-

point improvement achieved in this work over previous reports. (f) Shelf stability of unencapsulated devices stored under $N_2$ conditions, showing the normalized PCE as a function of storage time. (g) Normalized thermal stability at 65°C measured in glove box for unencapsulated devices.

To assess the robustness of the EDA passivation across different processing solvents, we tested anisole, chlorobenzene, toluene, diethyl ether and ethyl acetate.[17] EDA consistently improved device performance, although the improvement was less pronounced with the more polar ethyl acetate (Fig. S26), likely because stronger solvent-perovskite interactions perturb the surface and compete with EDA-induced restructuring.

To further examine the generality of this surface chemistry, we considered recent literature reports and further investigated passivation molecules with varying carbon chain lengths (Fig. S27). Interestingly, despite their closely related molecular structures, short-chain diamines (2C) consistently outperformed their longer-chain analogues (3–5C). The advantage of short-chain neutral amines can be attributed to the higher strength of diamine chelation in removing excess surface Sn and promoting surface restructuring. In the case of the ammonium salts, the superior performance of short-chain ligands could be associated with reduced insulating barrier formation upon interaction with the perovskite surface.[43,44,45] In addition, we further verified that, for longer molecules, neutral diamines are also more effective than their corresponding diammonium counterparts in passivating Sn-Pb perovskite surfaces and improving device performance (Fig. S28), probably owing to the stronger chelating effect towards excess Sn and surface restructuring as discussed above.

Finally, we investigated the long-term stability of our Sn-Pb devices without encapsulation. MA-free Sn-Pb perovskites are known for their favorable thermal stability.[46] Notably, compared with the reference devices, which retain 93% of their initial PCE after 215 days of storage, both the EDA- and $EDAI_2$-passivated devices exhibited markedly enhanced shelf-life stability, fully preserving their initial PCE under inert-atmosphere conditions (Fig. 4f). Interestingly, the $EDAI_2$-treated devices showed a gradual increase in PCE during the early stage of storage, suggesting a time-dependent interfacial passivation process that progressively eliminates residual defect states and further optimizes the interface. This behavior highlights the fundamentally different interfacial evolution pathways induced by the neutral and protonated forms of passivation studied here. Under continuous AM 1.5G illumination, maximum power point (MPP) tracking revealed a rapid decay for the reference devices, indicative of pronounced interfacial degradation. In contrast, both devices built with EDA-and $EDAI_2$-treated films showed significantly enhanced operational stability, with the former retaining 80% of their initial performance after 200 h (Fig. S29). Thermal stability measurements at 65 °C further revealed a pronounced benefit of EDA passivation. Devices based on EDA-treated films retained over 90% of their initial PCE after 600 h of continuous heating, whereas $EDAI_2$-treated and untreated devices retained only 63% and 60%, respectively. The improved operational and thermal stability is attributed to EDA-induced surface restructuring and improved surface stoichiometry, which stabilize the perovskite/C60 interface and mitigate interfacial degradation. Collectively, these results establish EDA-based surface engineering as an effective strategy for simultaneously improving the efficiency and long-term stability of Sn-Pb perovskite photovoltaics.

In summary, we demonstrate that molecular passivation in Sn-Pb perovskites is governed by the underlying surface chemistry and stoichiometry. Although EDA and $EDAI_2$ share the same molecular backbone, they initiate distinct interfacial processes involving direct metal chelation, surface reconstruction and defect-coupled proton transfer facilitated by iodide migration. These pathways determine whether molecular treatment predominantly passivates electronic defects or additionally reconstructs the surface composition and energetics. Combined with Br-2PACz-based buried-interface modification, EDA-induced surface reconstruction enables methylammonium-free Sn-Pb perovskite solar cells with a PCE of 25.86%, retaining their initial performance after 215 days of storage under an inert atmosphere and more than 90% of their initial PCE after 600 h at 65 °C. More broadly, our results demonstrate that the effectiveness of molecular passivation depends on the intrinsic defect chemistry of the perovskite surface, which in turn is shaped by composition, stoichiometry and processing conditions. This mechanistic framework helps reconcile the varying effects of chemically related passivation treatments reported in the literature and provides a basis for the rational design of molecular interfaces in metal halide perovskites.

**Acknowledgements:**
This work is financially supported by Horizon Europe Framework Programme (HORIZON) research and innovation programme under the grant agreement No 101084124 (Diamond). This work is part of the SolarNL program, which is supported by the Dutch National Growth Fund (Nationaal Groeifonds). Part of the work is also supported by the European Union (ERC-Advanced Grant, DEOM, 101055097). Views and opinions expressed are, however, those of the author(s) only and do not necessarily reflect those of the European Union or the European Research Council. Neither the European Union nor the granting authority can be held responsible for them.

**Author contributions:**
M.A.L. conceived the study, supervised the project and organized the resources. L.C. designed and conducted most of the experimental work and performed the data analysis. M.A.L., L.C., D.G.R. and L.P. contributed to the design of specific experiments and the interpretation of the results. M.P., S.Z., G.N. and D.G.R. contributed to device fabrication and data analysis. M.A. and G.P. performed the GIWAXS measurements and the related analysis. S.L., E.M. and F.D.A. performed the simulations. D.G.R. conducted the AFM and KPFM measurements. J.K. and L.P. performed the NMR measurements and analysis. D.V.N., A.G.C. and P.R. performed the XPS measurements and analysis. J.L. and J.P. performed the SEM measurements. A.S. performed the PL and TRPL measurements. L.C. wrote the initial draft of the manuscript. All authors discussed the results and contributed to revising the manuscript.

**Competing interests:**
The authors declare no competing interests.

**Data and materials availability:**
All data needed to evaluate the conclusions in the paper are present in the paper or the supplementary materials.

**Supplementary Materials**

# Controlling Surface Chemistry in Tin-Lead Perovskites for High-Efficiency Solar Cells

Lijun Chen[1], Sitong Zhao[1], Gaukhar Nigmetova[1], David Garcia Romero[1], Matteo Pitaro[1], Sofia Lerda[2,3], Edoardo Mosconi[3], Michael Aidi[1], Douwe van Netten[1], Julia Kraft[1], Jiaxiong Li[1], Jacopo Pinna[1], Asya Serafimova[1], Antonija Grubišić-Čabo[1], Loredana Protesescu[1], Giuseppe Portale[1], Petra Rudolf[1], Filippo De Angelis[2,3,4], Maria Antonietta Loi[1]*

1.Zernike Institute for Advanced Materials, University of Groningen, Nijenborgh 3, 9747 AG Groningen, The Netherlands
2. Department of Chemistry, Biology and Biotechnology, University of Perugia, INSTM, Via Elce di Sotto 8, Perugia 06123, Italy
3. Computational Laboratory for Hybrid/Organic Photovoltaics (CLHYO), Istituto CNR di Scienze e Tecnologie Chimiche "Giulio Natta" (CNR-SCITEC), Via Elce di Sotto 8, Perugia 06123, Italy
4. SKKU Institute of Energy Science and Technology (SIEST), Sungkyunkwan University (SKKU), Suwon 16419, Republic of Korea
*Corresponding author. Email: m.a.loi@rug.nl

**The PDF file includes:**
Materials and Methods
Supplementary Texts 1-6
Supplementary Figures 1-29
Supplementary Tables 1-8
Supplementary References 1-34

## Methods

**Materials:** Formamidinium iodide (FAI, 98%, TCI), Cesium Iodide (CsI, 99.999%, Sigma Aldrich), lead iodide ($PbI_2$, 99.999%, Thermo Fisher Scientific), tin (II) iodide ($SnI_2$, 99.99%, Sigma Aldrich), tin (II) fluoride ($SnF_2$, 99%, Sigma Aldrich), N,N-dimethylformamide (DMF, 99.8%, Thermo Fisher Scientific), dimethyl sulfoxide (DMSO, 99.8%, Thermo Fisher Scientific), anisole (99.7%, Sigma Aldrich), ethylenediamine (EDA, 99%, Sigma Aldrich), ethylenediammonium diiodide ($EDAI_2$, 98%, Sigma Aldrich), buckminsterfullerene ($C_{60}$, 99.99%, Sigma Aldrich), bathocuproine (BCP, 99.99%, TCI), isopropyl alcohol (IPA, 99%, Sigma Aldrich), [2-(3,6-Dibromo-9H-carbazol-9-yl) ethyl] phosphonic acid (Br-2PACz, Dyenamo), Ethanol (EtOH, 99.5%, Thermo Fisher Scientific), Toluene (99.8%, Sigma Aldrich), diethyl ether (99.8%, Sigma Aldrich), chlorobenzene (99.8%, Thermo Fisher Scientific), ethyl acetate (99.7%, Sigma Aldrich), 1,2-diaminopropane (DAP, 99%, Sigma Aldrich), 1,4-butanediamine (BDA, 99%, Sigma Aldrich), and 1,3-Propanediamine (PrDA, 98%, TCI), 1,4-butanediamine dihydriodide ($BDAI_2$, Sigma Aldrich), 1,3-Propanediamine dihydriodide ($PrDAI_2$, 98%, Sigma Aldrich), and 1,5-pentanediamine (PeDA, 98%, TCI).

**Precursor solution preparation:** The SAM solution was prepared by dissolving Br-2PACz in EtOH at a concentration of 0.1 mM. The Sn-Pb perovskite solution was prepared by dissolving CsI (0.425 M), FAI (1.275 M), $PbI_2$ (0.85 M), $SnI_2$ (0.85 M), and $SnF_2$ (0.085 M) in a mixed solvent of 0.8 mL of DMF and 0.2 mL of DMSO (DMF: DMSO = 4:1). The solution was stirred overnight at room temperature and subsequently filtered through a 0.22 µm PTFE filter before use. For post-treatment solutions, the EDA solution was prepared by diluting raw EDA liquid in IPA or other solvents to a final concentration of 0.1 mM. The $EDAI_2$ solution was prepared by dissolving 0.75 mg $EDAI_2$ powder in IPA and stirring the mixture overnight at 50 °C. Other passivation solutions were prepared in IPA at various concentrations, followed by stirring overnight at room temperature.

**Solar cell fabrication:** ITO glass substrates (3 x3 $cm^2$) were sequentially cleaned in detergent, deionized water, acetone, and isopropanol (10 min each in an ultrasonic bath), followed by 30 min of UV-ozone treatment in air. The substrates were then immersed overnight in the EtOH/Br-2PACz solution, spin-dried at 3000 rpm for 30 s, and annealed at 100 °C for 2 min. Subsequently, they were rinsed with EtOH during a second 30 s spin at 3000 rpm to remove any aggregated Br-2PACz, followed by a final annealing step at 100 °C for 1 min in air. Next, the substrates were transferred into a glovebox, and the perovskite solution was dispensed onto the substrate and spin-coated at 4000 rpm for 80 s. At 25 s after the start of spin-coating, 250 µL of anisole was dropped onto the spinning substrate. The resulting perovskite films were then annealed at 100 °C for 10 min. For surface treatment, EDA or other passivation solutions at varying concentrations were applied by allowing the solution to stand on the perovskite surface for 1 min, followed by spin-coating at 3000 rpm for 30 s and annealing at 100 °C for 1 min. After passivation, the samples were transferred to a vacuum chamber for thermal deposition, where 30 nm of $C_{60}$, 6 nm of BCP, and 100 nm of Ag were sequentially deposited.

**Solar cells characterization: J-V curves** were measured with a Keithley 2400 source meter at 295 K under an irradiation intensity of 100 mW/cm$^2$, calibrated with a reference Si solar cell from Newport. The scan rate is 100 mV/s with a voltage step of 10 mV. An aperture shade mask with an area of 0.04 cm$^2$ was used to define the active area.

**EQE** measurements were performed using a setup consisting of a xenon lamp and three filter wheels for sample illumination. The photon flux was calibrated with Newport optical power detectors (818-SL and 818-IR), and the resulting photocurrent was measured using a lock-in amplifier (Stanford Research Systems SR830 DSP). A spectral mismatch factor was applied to correct for differences between the spectrum of the lamp used during EQE measurements and the AM1.5G reference spectrum.

**SEM** images were recorded by NovaNano SEM 650 microscope with an acceleration voltage of 10 kV.

**AFM** measurements were carried out in air using a Bruker Multimode 8 microscope operated in ScanAsyst mode.

**KPFM** measurements were performed on the same instrument using SCM-PIT-V2 probes (resonant frequency, 75 kHz; spring constant, 3 N m$^{-1}$) in Electrical & Magnetic Lift Mode with surface potential (FM-KPFM).

**FTIR** spectra of the films and solvents were acquired in the range of 500-4000 cm$^{-1}$ using a Shimadzu IRTracer-100 spectrometer inside a glovebox.

**GIWAXS** measurements were acquired using the Multipurpose X-ray Instrument for Nanostructure Analysis (MINA). The instrument is built around a Cu rotating anode source (λ=1.5413 Å). GIWAXS images were captured using a Bruker Vantec 500 detector equipped with a 2048 x2048 pixel layout and pixel dimension of 68 x68 µm placed 93 mm away from the sample and using incident angles of 0.4° and 4° were used to collect surface and bulk structural information, respectively. A single frame exposure lasting 1800 s was recorded at each incident angle under vacuum conditions (~10$^{-3}$ mbar).

**GISAXS** measurements were captured with samples placed in a $N_2$ atmosphere with a Pilatus 300K detector equipped with a 487 x618 pixel layout and pixel dimension of 172 x172 µm using the same X-ray source. The sample-to-detector distance was ~3120 mm and the incident angles were 0.4°, 0.5°, and 0.6°. A single frame exposure lasting 3600 s was recorded at each incident angle. Both GIWAXS and GISAXS patterns were elaborated using the GIXGUI Matlab suite. The covered angular range was calibrated using the position from known diffraction peaks of standard silver behenate and $Al_2O_3$ samples for GISAXS and GIWAXS, respectively.
**NMR** spectra were recorded on a Bruker Advance Neo 600 (600 MHz) using a sealed coaxial DMSO-d6 capillary for field locking and shimming.
**PL and TRPL** measurements were performed using a Ti:Sapphire laser (Coherent, Mira 900), pumped by a 532 nm Nd:YAG laser. The Ti:Sapphire laser generates 150 fs pulses at 800 nm with a repetition rate of 76 MHz. The excitation laser was frequency-doubled to a 400 nm beam using a second harmonic generator. Steady state PL was generated using a 20-fold pulse-picked excitation at 5 µW focused down to a 150 µm spot and collected into a spectrograph with a 30 l/mm grating and a spectrally calibrated infrared sensor (1.7 µm InGaAs, Andor iDus). TRPL measurements employed a 640-fold pulse picker excitation power at 5 µW and were recorded using a Hamamatsu Streak camera in Single-Sweep mode. In all cases, the samples were measured in a nitrogen atmosphere. All measurements were carried out in reflection geometry under the same conditions.
**XPS** measurements of the samples were transferred from the glovebox to the spectrometer in a sealed plastic bag filled with nitrogen and introduced into the loadlock under nitrogen atmosphere. The samples were subsequently analyzed under ultrahigh-vacuum conditions ($<5 \times 10^{-9}$ mbar) using a Scienta R4000 hemispherical analyzer at normal emission. The X-rays illuminated the whole width of the sample; the overall experimental resolution was 0.4 eV and binding energies were referenced to the C-C/C-H component of the C 1s core-level spectrum at 284.8 eV. One sample was analyzed for each surface treatment, and two spots were measured on each sample to check for homogeneity. Spectra analysis was done using the least squares curve fitting program WinSpec developed at the LISE, University of Namur, Belgium, and included a Shirley baseline subtraction and a peak deconvolution using a linear combination of Gaussian and Lorentzian functions, taking into account the experimental resolution. The spectra were fitted with a minimum number of peaks consistent with the structure of the surface. Binding energies of isolated peaks are given at ±0.05 eV; when more than one component was needed to reproduce the raw data, the error in the component position was ±0.1 eV. The uncertainties in the intensity determinations were approximately 1%.
**XRD** measurements were performed using the Bruker D8 Advanced diffractometer with Cu-Kα radiation (λ =1.5418 Å) in standard Bragg-Brentano geometry scanning a 2θ range from 5-45° with a step size of 0.02°/sec. The measurements were carried out in atmospheric conditions at room temperature (T~293K). Phase identification and structure refinement were carried out with Profex 5.6.1 making use of the COD and ICSD database.
**EIS** measurements were performed with a Solartron 1260 impedance gain-phase analyzer. And the DC voltage was set to open-circuit conditions over a frequency range of 0.1 Hz to 1 MHz under a $N_2$ atmosphere.
**UV-vis** absorption spectra were measured using a Shimadzu UV-Vis-NIR spectrophotometer (UV-3600) under a $N_2$ atmosphere.

**MPP** tracking measurements were conducted using the same setup and solar simulator as those employed for the J-V characterization, with a 400 nm UV filter inserted during the measurements.

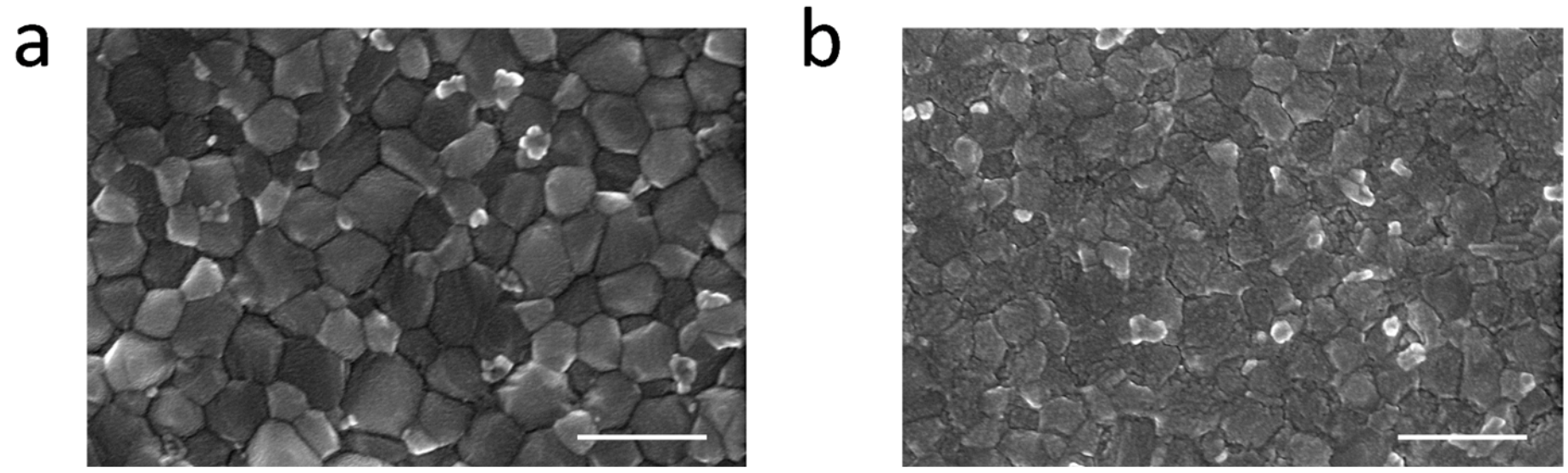

**Supplementary Fig. 1** SEM images of reference perovskite films before (a) and after (b) IPA treatment. Scale bar: 1 µm.

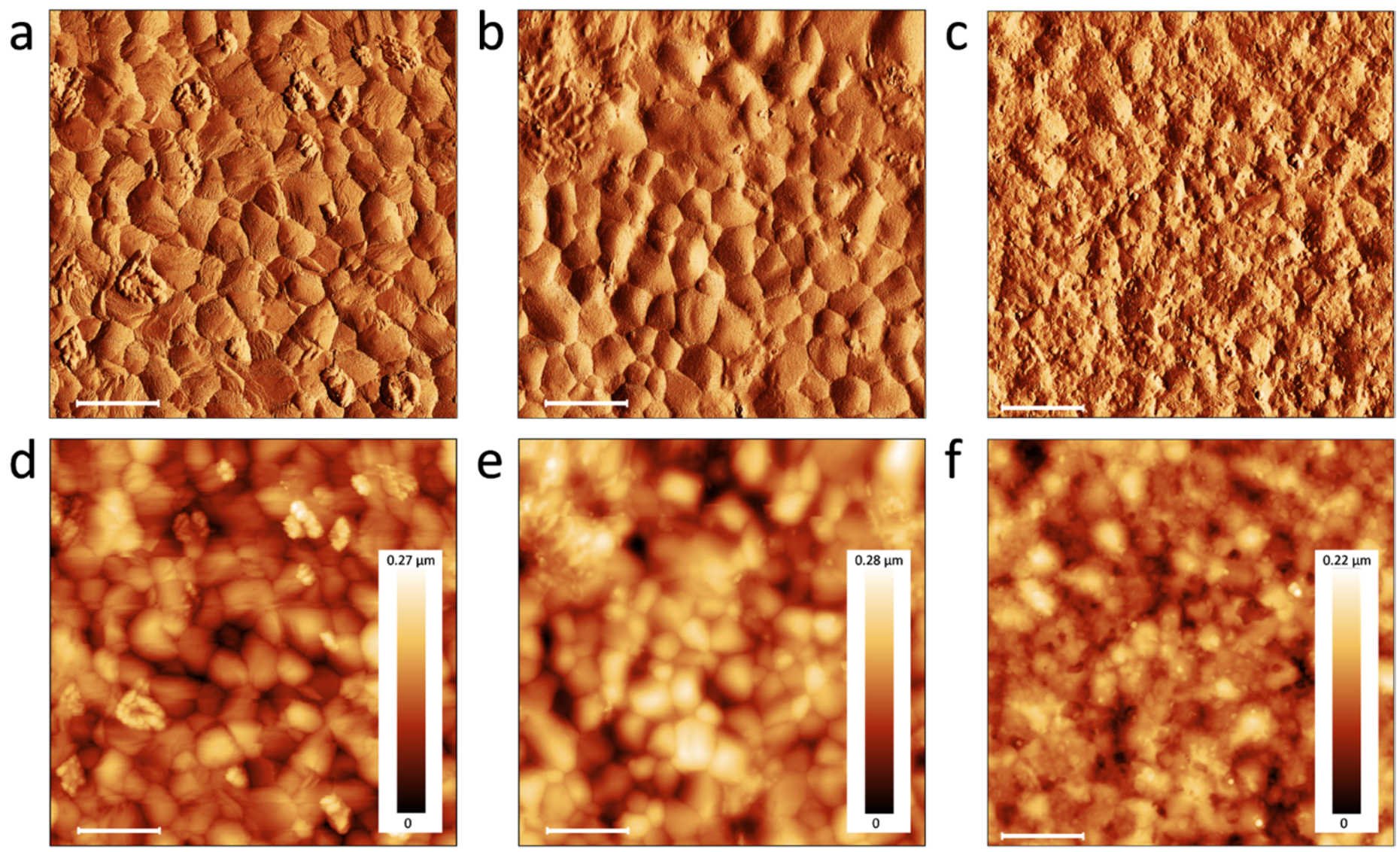

**Supplementary Fig. 2** Peak-force error images of the reference (a), $EDAI_2$-treated (b), and EDA-treated (c) Sn-Pb perovskite films. (d-f) Corresponding AFM topographic images. Scale bar: 1 µm.

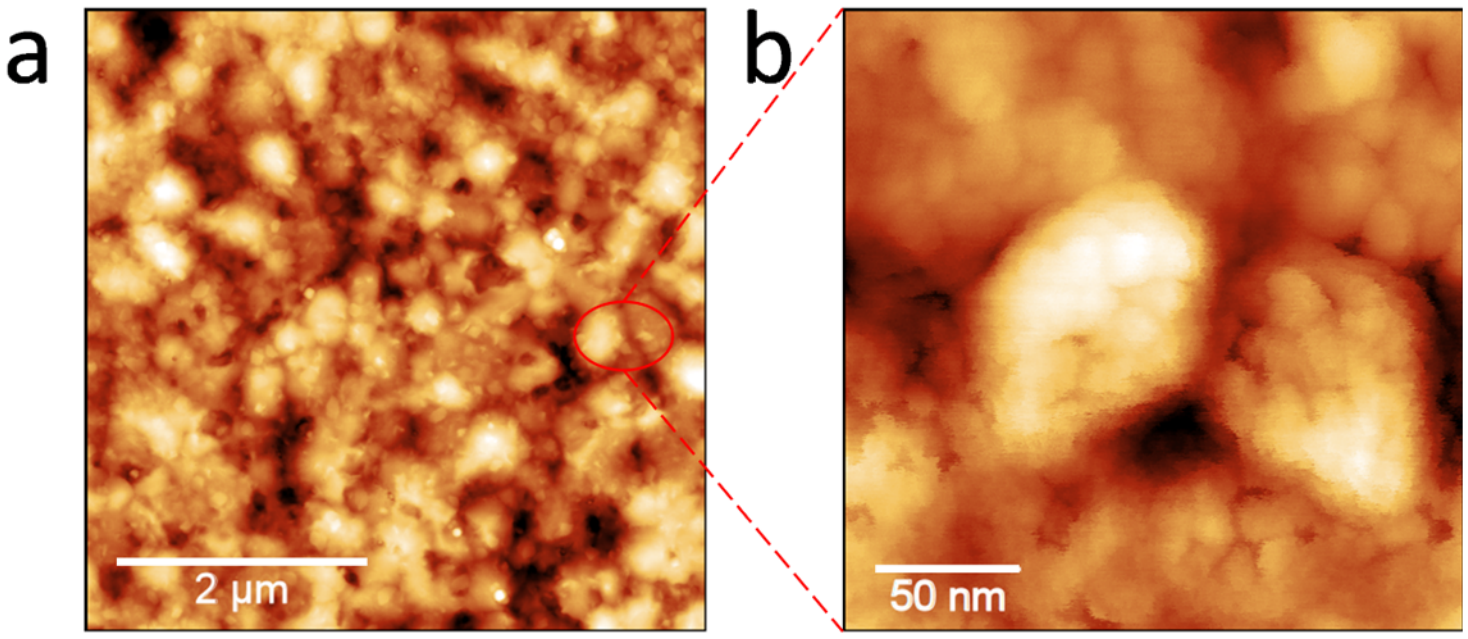

**Supplementary Fig. 3** (a) AFM topographic image of a Sn-Pb perovskite film treated with EDA. (b) High resolution AFM image highlighting the nanostructures.

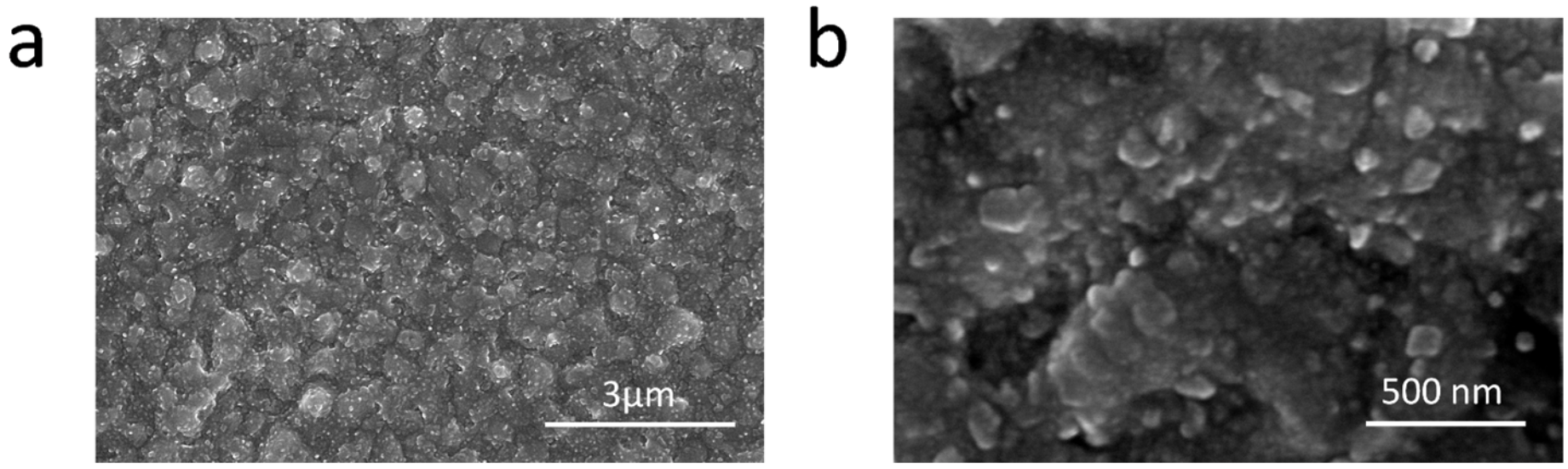

**Supplementary Fig. 4** (a) SEM topographic image of a Sn-Pb perovskite film treated with EDA. (b) High resolution SEM image highlighting the nanostructures.

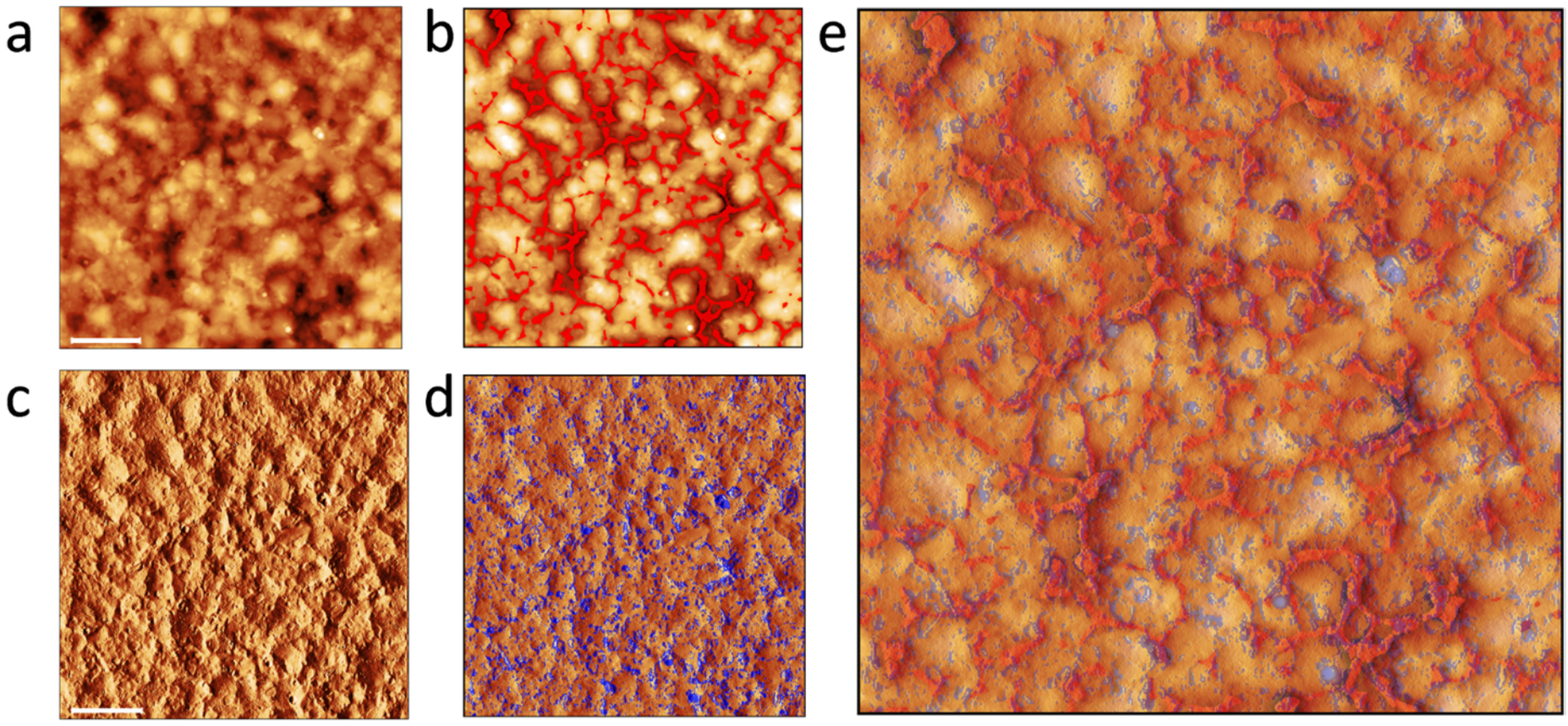

**Supplementary Fig. 5** Correlated height and peak-force error images of EDA-treated Sn-Pb perovskite films. (a, b) AFM height images and (c, d) corresponding peak-force error images. Scale bar: 1 µm. (e) A combined high-resolution topography image.

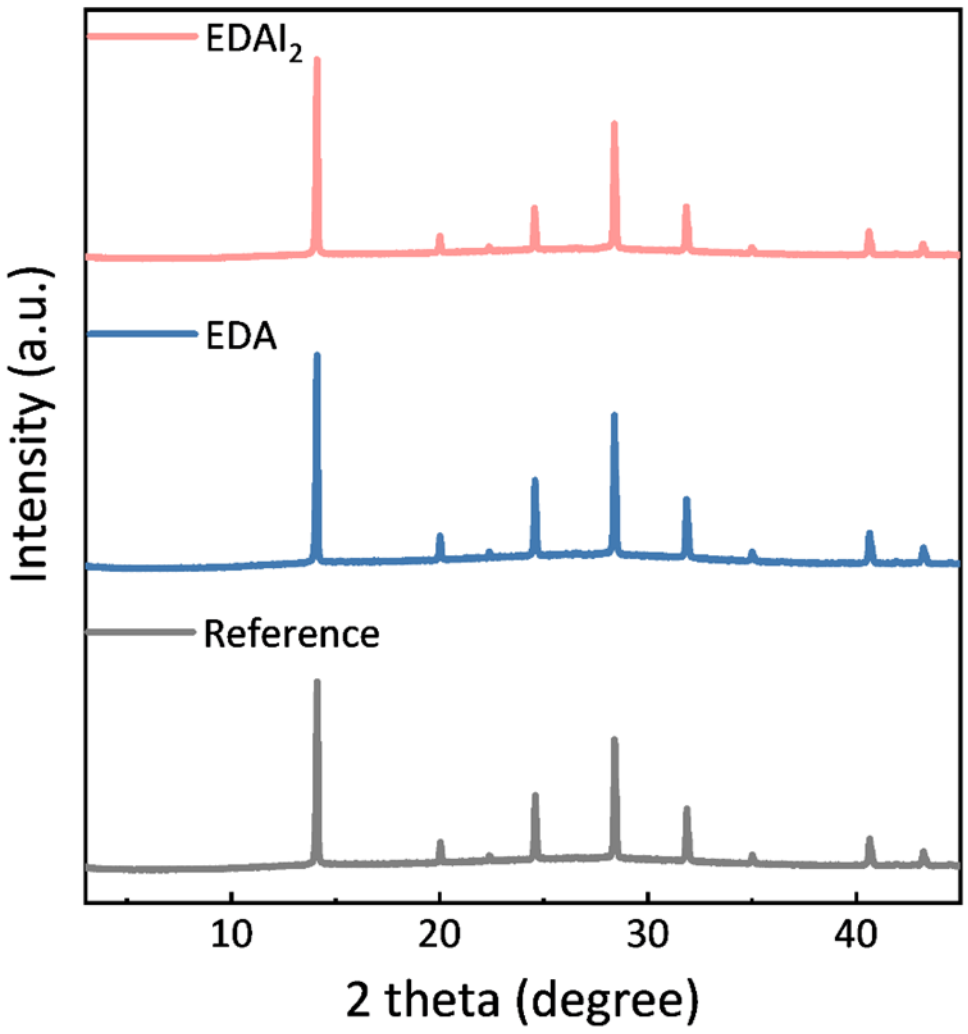

**Supplementary Fig. 6** XRD patterns of Sn-Pb perovskite films of reference, EDA-treated, and $EDAI_2$-treated samples.

**Supplementary Text 1**

Distinct Bragg reflections were observed for all samples, with no appreciable differences between their diffraction patterns. Rietveld refinement was initially performed using a tetragonal structure with the P4/mbm space group (no. 127), based on the model reported by Toney et al.[1] for $Cs_{0.5}FA_{0.5}PbI_3$ but modified to incorporate Pb/Sn site occupancies of 50:50 and the appropriate Cs/FA ratio. For the untreated reference sample, the refined lattice parameters were a = b = 8.8746 ± 0.0007 Å and c = 6.2914 ± 0.0006 Å, with similar values obtained for the passivated samples (Fig. S7 and Table S1). The model reproduced the positions and number of reflections well, although discrepancies in their relative intensities could not be resolved by refining the site occupancies, atomic coordinates or Debye–Waller factors. Introducing preferred-orientation corrections improved the fit only when sixth- to tenth-order spherical harmonics were used. Because such high-order terms may overfit the data and lead to an erroneous assignment of texture, they were not considered reliable.

We therefore additionally performed composition-independent Le Bail refinements to determine the crystal system, space group and unit-cell parameters without relying on specific atomic identities, occupancies or coordinates.[2] Although both cubic Pm-3m and tetragonal P4mbm models could be applied to the data, the cubic model provided a poorer fit and failed to account for several reflections (Fig. S8). By contrast, the tetragonal model indexed all observed reflections and yielded substantially better profile agreement, with lattice parameters consistent with those obtained from Rietveld refinement (Table S2). These results support assignment of the tetragonal P4/mbm structure to both untreated and passivated samples. Previous studies of Pb-only perovskites with similar compositions have likewise shown that partial substitution of $FA^+$ with $Cs^+$ stabilizes the tetragonal phase at room temperature,[3,4] whereas the corresponding Sn-only system reportedly remains cubic upon low-level $Cs^+$ substitution.[5]

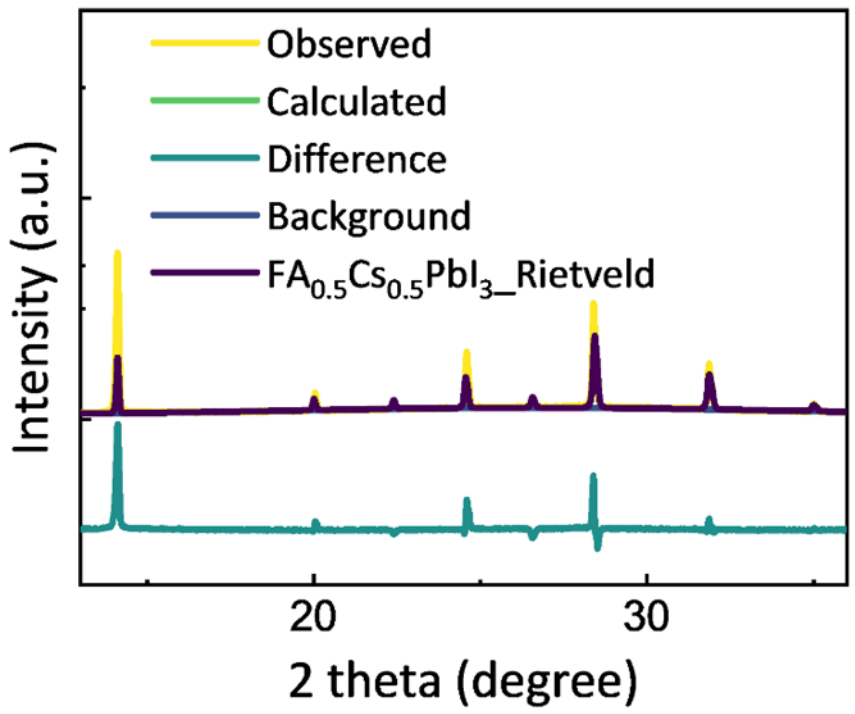

**Supplementary Fig. 7** XRD Rietveld refinement with the tetragonal phase.

**Supplementary Table 1** Refined unit cell parameters of the Sn-Pb perovskite using Rietveld refinement.

| | Reference | EDA-treated | $EDAI_2$-treated |
|---|---|---|---|
| System | Tetragonal | Tetragonal | Tetragonal |
| Space Group | P4/mbm | P4/mbm | P4/mbm |
| a (Å) | 8.87078 ± 0.000096 | 8.8772 ± 0.00021 | 8.8852 ± 0.00026 |
| b (Å) | - | - | - |
| c (Å) | 6.29066 ± 0.000081 | 6.2775 ± 0. 00034 | 6.2851 ± 0. 00025 |
| Unit Cell Volume ($Å^3$) | 495.02 | 494.70 | 496.19 |

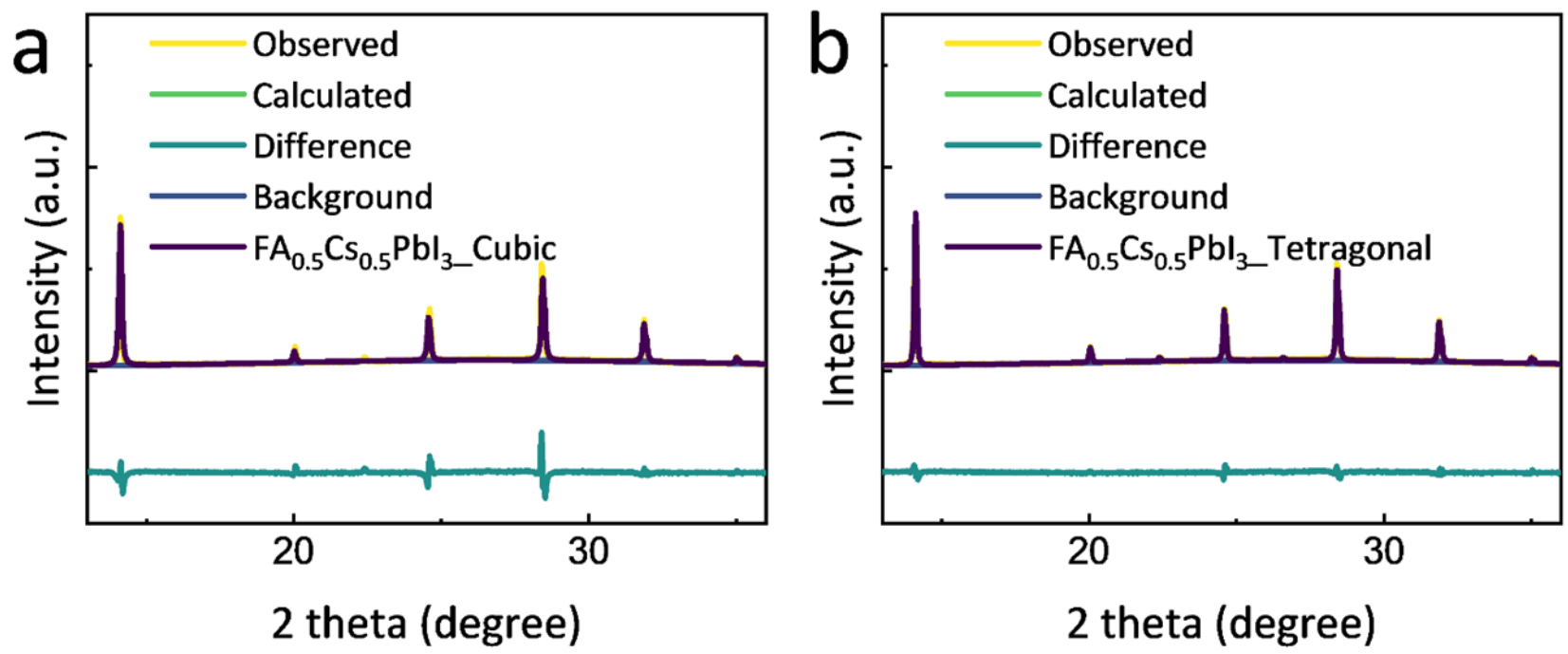

**Supplementary Fig. 8** Le Bail refinement for (a) cubic and (b) tetragonal structure.

**Supplementary Table 2** Refinement unit cell parameters of the Sn-Pb perovskite using the Le Bail method.

| | Reference | EDA-treated | $EDAI_2$-treated |
|---|---|---|---|
| System | Tetragonal | Tetragonal | Tetragonal |
| Space Group | P4/mbm | P4/mbm | P4/mbm |
| a (Å) | 8.89645 ± 0.000042 | 8.89530 ± 0.000039 | 8.89876 ± 0.000028 |
| b (Å) | - | - | - |
| c (Å) | 6.26467 ± 0.000041 | 6.26244 ± 0.000037 | 6.27491 ± 0.000029 |
| Unit Cell Volume ($Å^3$) | 495.83 | 495.52 | 496.90 |

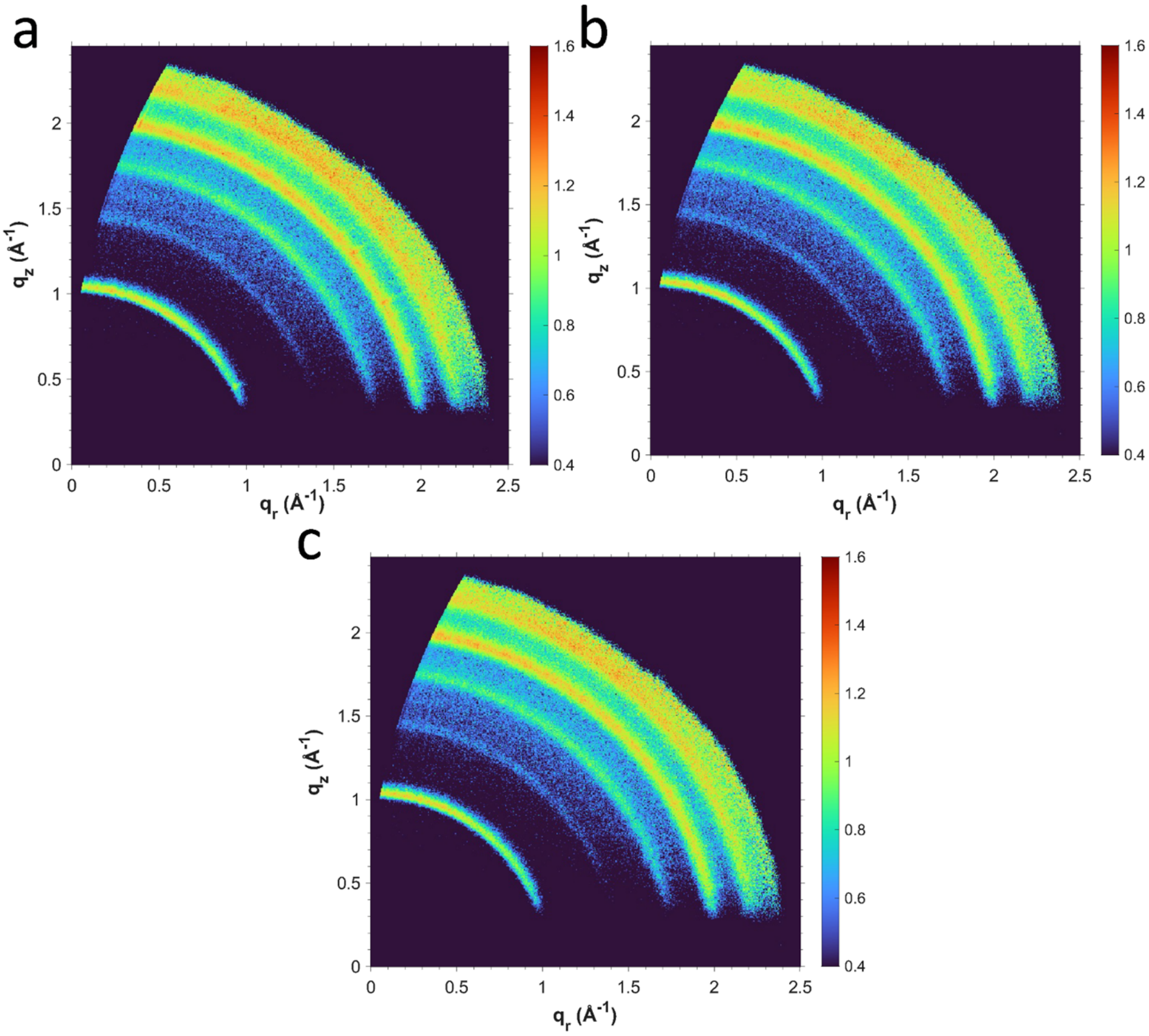

**Supplementary Fig. 9** GIWAXS patterns of (a) reference, (b) $EDAI_2$-treated and (c) EDA-treated Sn-Pb perovskite films, acquired using an incident angle of 4°.

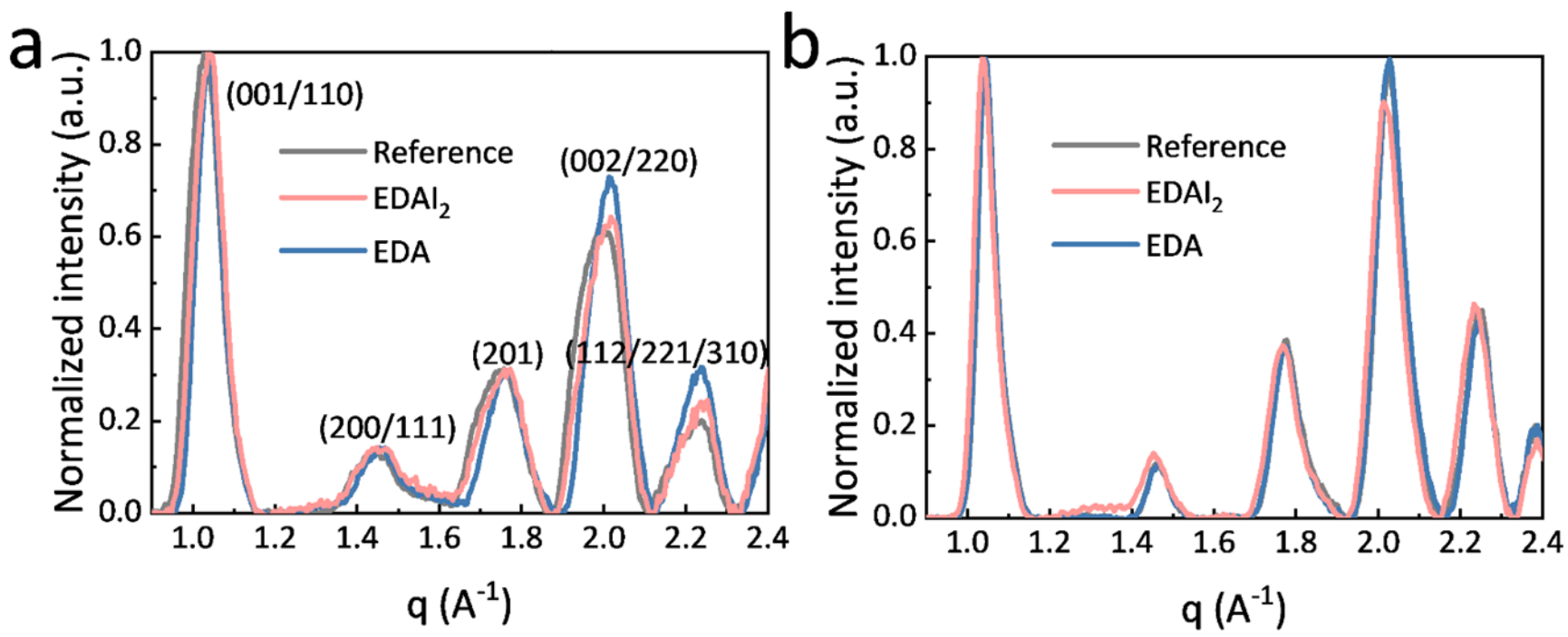

**Supplementary Fig. 10** Comparison of the average radial GIWAXS intensity profiles I(q) vs q for (a) 0.4° and (b) 4° incident angles.

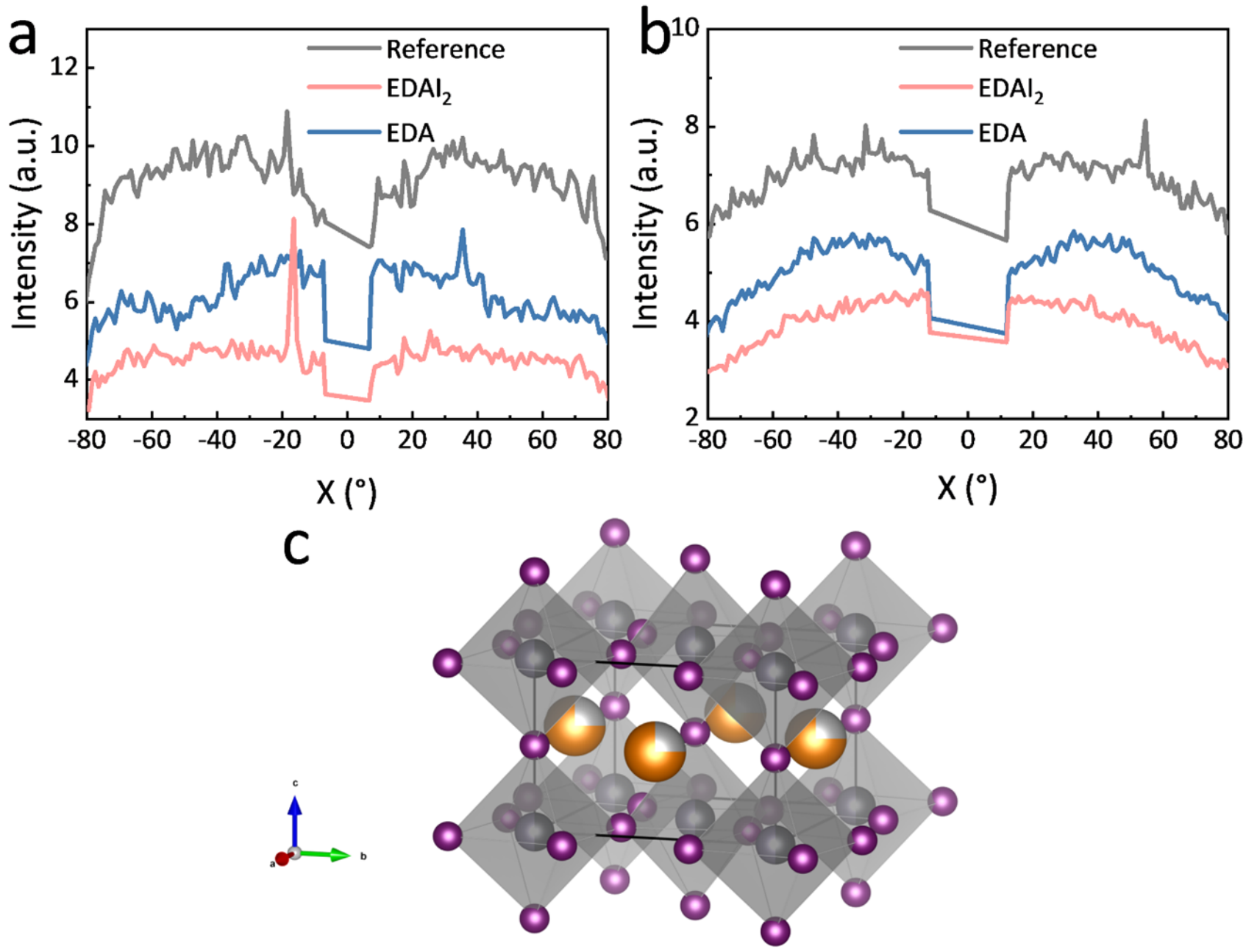

**Supplementary Fig. 11** Comparison of the azimuthal GIWAXS profiles for the (a) 001/110 and (b) 201 diffraction rings. (c) Structure for the small but measurable preferential orientation of the unit cells in the EDA-treated Sn-Pb perovskite film.

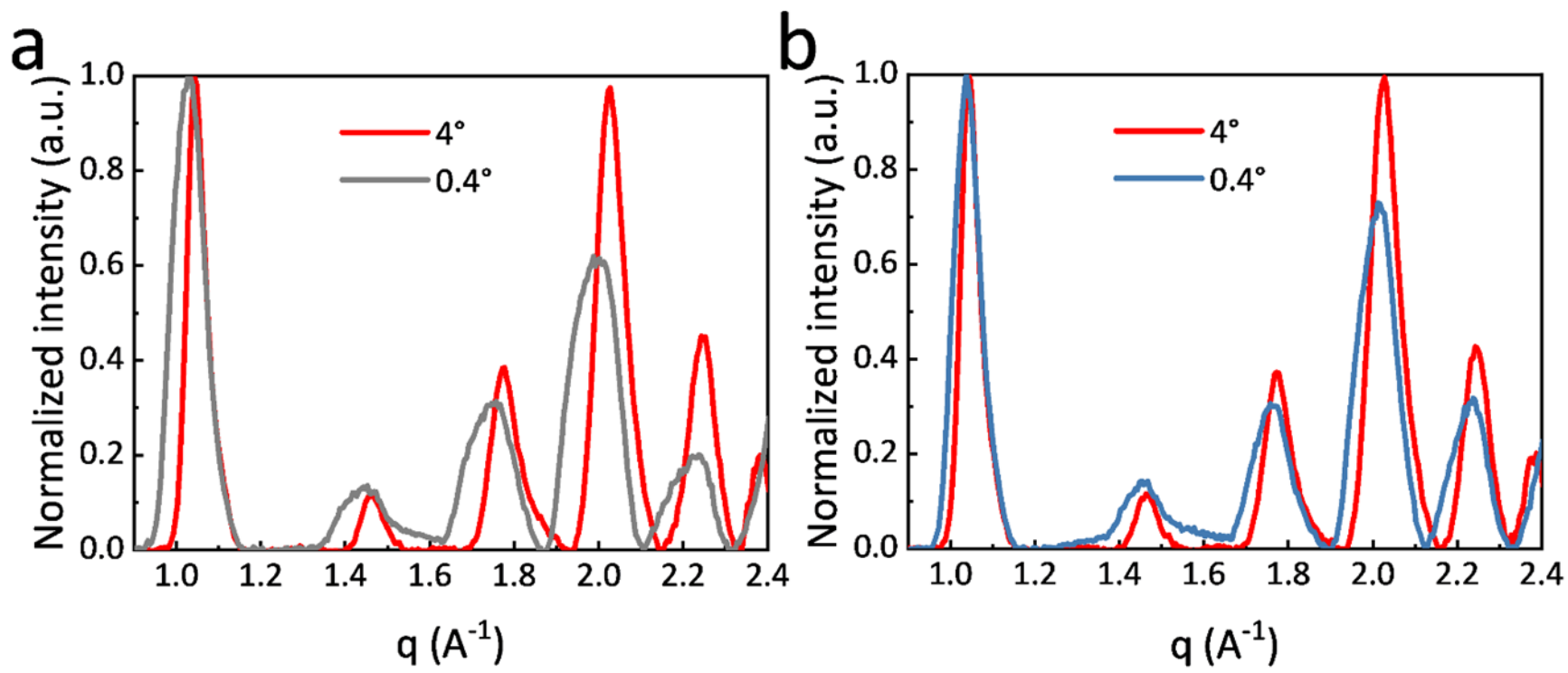

**Supplementary Fig. 12** Comparison of the average radial GIWAXS profiles acquired at incident angle 0.4° (surface) and 4° (bulk) for (a) reference and (b) EDA-treated Sn-Pb perovskite films.

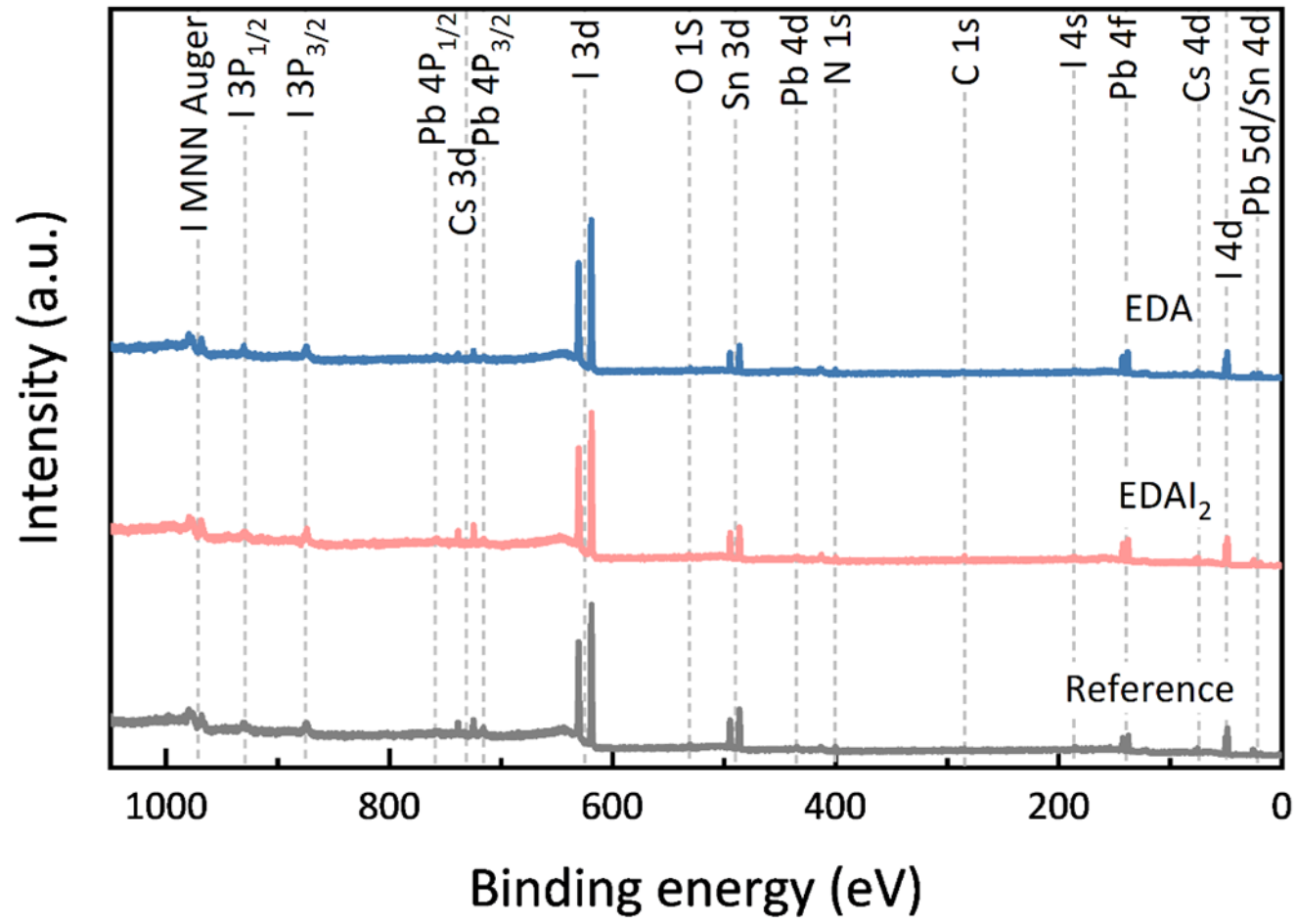

**Supplementary Fig. 13** XPS survey spectra of reference, EDA-treated, and $EDAI_2$-treated Sn-Pb perovskite films. The spectra for the three films are offset in intensity for clarity, and each observed core level is labelled accordingly.

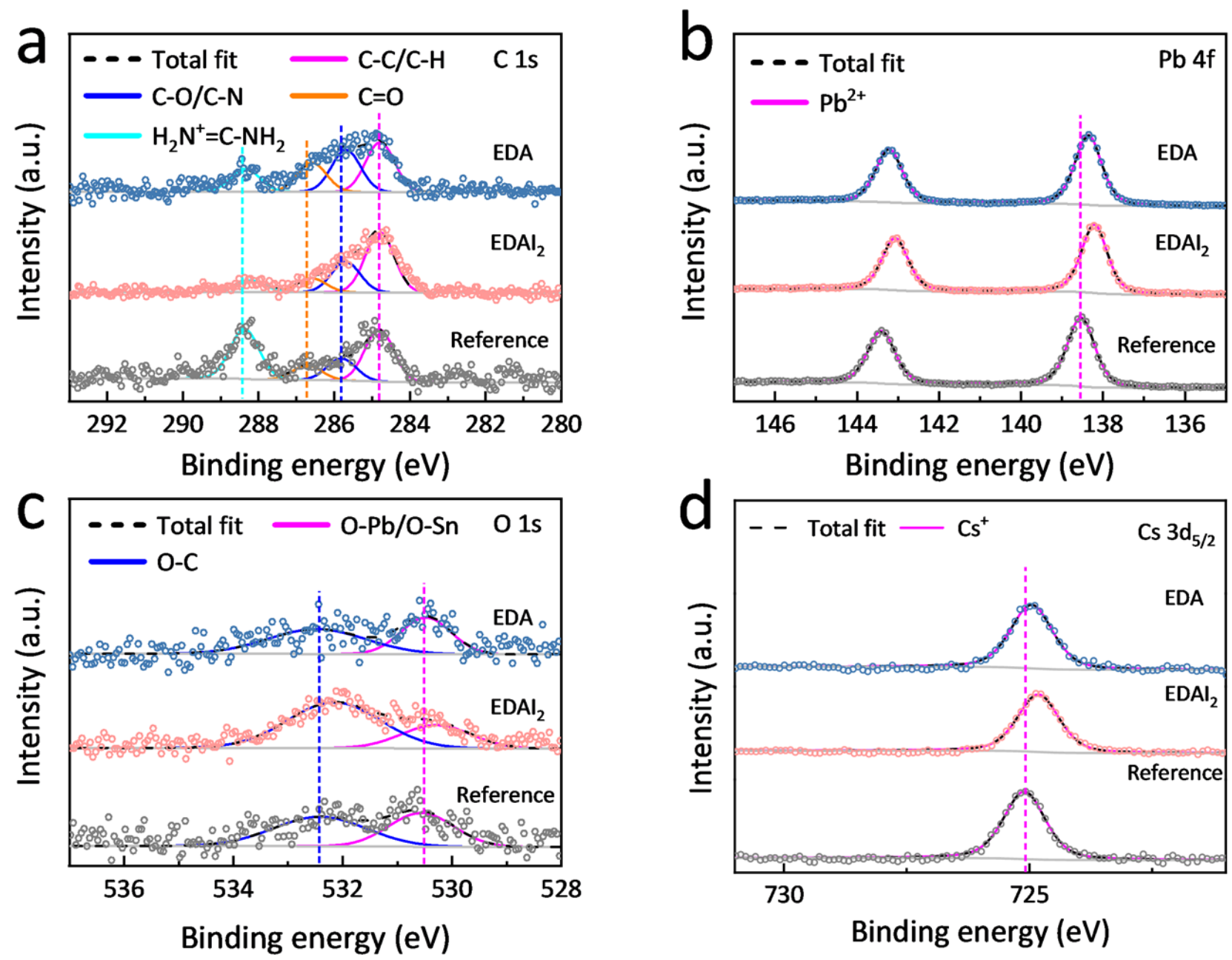

**Supplementary Fig. 14** XPS spectra of the (a) C 1*s*, (b) Pb 4*f*, (c) O 1*s* and (d) Cs 3*d* core-level regions (dots) of the reference, $EDAI_2$-treated, and EDA-treated Sn-Pb perovskite films and corresponding fits (solid lines). The spectra are vertically offset for clarity, while the vertical dashed lines indicate changes in the binding energy

positions of the fitted components between the different surfaces.

**Supplementary Text 2**

The XPS spectra of the N 1*s*, Cs 3*d*, Sn 3*d*, I 3*d*, C 1*s*, Pb 4*f* and O 1*s* core-level regions of reference, $EDAI_2$-treated and EDA-treated films are presented in Fig. 2 and Fig. S14.

The most pronounced changes between the reference, $EDAI_2$-treated, and EDA-treated surfaces are observed in the N 1*s* core level spectra, as shown in Fig. 2a. For the reference sample surface, the N 1*s* spectrum consists of a single component at BE=400.6 eV, which is assigned to the nitrogen atoms in the $FA^+$ molecule ($H_2N^+$=C-$NH_2$).[6,7] Following $EDAI_2$ or EDA treatment, an additional component appears at BE=399.9 eV. This component is assigned to the nitrogen atoms of the EDA molecule (R–$NH_2$), demonstrating the presence of EDA at the treated surfaces.[6] The spectra were also examined for evidence of protonated EDA (R-$NH_3^+$), which is expected[8] at approximately 402.1 eV, as indicated by the yellow dashed line. However, no additional intensity is observed at this binding energy, indicating that there is no protonated EDA present on either treated sample surface.

The Sn 3*d* core-level spectra in Fig. 2b contains two components. The dominant component at BE=486.8 eV is assigned to $Sn^{2+}$ in the perovskite lattice,[6,8,9] while the less intense component at 487.6 eV is assigned to $Sn^{4+}$ associated with surface oxidation.[8,9] Quantification of the two components (Tables S3 and S4) demonstrates that the $Sn^{2+}$ is the predominant species in all three samples, whereas the contribution from $Sn^{4+}$ remains minimal. The other peaks at BE=495.1 and 495.9 eV are attributed to the $Sn^{2+}$ and $Sn^{4+}$ components in the Sn $3d_{3/2}$ core level.

The I 3*d* core-level spectra is presented in Fig. 2c and are fitted with a single component at 619.4 eV, which is assigned to $I^-$ in the perovskite lattice.[6,7,8,9] The position of the $I^-$ component shifts towards lower binding energy following $EDAI_2$ or EDA treatment. The magnitude and direction of the binding energy shifts are comparable to those observed for the Pb 4*f* core level. The I $3d_{3/2}$ component is observed at BE = 630.9 eV.

The C 1s spectra (Fig. S14a) are fitted with four components. The most intense component at BE= 284.8 eV is assigned to C–C/C–H bonds and is used as the binding energy reference.[6] The components at 285.8 and 286.9 eV are attributed to C–O/C–N and C=O bonds, respectively,[8] where the C-N contribution is associated with EDA, while the oxygen-containing contributions are attributed to minor surface contamination introduced during sample transfer. A fourth component at BE=288.2 eV (light blue) is assigned to the carbon atom in $FA^+$ ($H_2N^+$=C-$NH_2$).[7] With the C-C/C-H component fixed at 284.8 eV, the other C 1*s* components change upon treatment. In particular, the C-N and the $FA^+$-related $H2N^+=C-NH_3$ components shift towards lower binding energy after $EDAI_2$ or EDA treatment, as indicated by the vertical dashed lines. These shifts indicate a modification of the local chemical environment and are consistent with the passivation-induced surface chemistry discussed in the main text, including the deprotonation of $EDAI_2$.

The Pb 4*f* core-level spectra of all three samples (Fig. S14b) are fitted with a single component at BE=138.5 eV, assigned to $Pb^{2+}$ in the perovskite lattice.[6,7,9] The $Pb^{2+}$ component shifts towards lower binding energy after treatment, by 0.3 eV for the $EDAI_2$-treated and 0.2 eV for the EDA-treated, respectively, relative to the reference sample. Similar shifts toward lower binding energies are observed for the Sn 3*d*, Cs 3*d* and I 3*d* core-level spectra, indicating a rigid shift of the core levels. The peak found at BE=143.3 eV is the Pb $4f_{5/2}$ core level.

The O 1*s* core-level spectra in Fig. S14c are fitted with two components. The component at BE=532.2 eV is assigned to O-C bonds associated with adventitious carbon,[8] while the component at 530.5 eV is assigned to oxygen bound to Pb or Sn (O-Pb/O-Sn).[6,8] The O 1*s* signal is weak for all three samples, with only a minor contribution from the O-Pb/O-Sn component relative to the perovskite constituents. Together with the minor $Sn^{4+}$ component observed in the Sn 3*d* core-level spectrum, these results indicate minimal surface oxidation in the reference, EDA-treated, and $EDAI_2$-treated films.

Fig. S14d shows the Cs $3d_{5/2}$ core-level spectrum of the three samples. The spectra are well fitted by a single component at BE= 725.1 eV, assigned to $Cs^+$ in the perovskite lattice.[7] Similar to the C-N and $H2N^+=C-NH_3$ components in the C 1*s* spectra, the $Cs^+$ component shifts towards lower binding energy after $EDAI_2$ or EDA treatment.

Further quantification of the XPS spectra is provided in Tables S3 and S4. Table S3 summarizes the atomic percentages obtained from the fitted peak areas of the core-level components relevant to the perovskite composition, namely $I^-$, $Pb^{2+}$, $Sn^{2+}$, $Sn^{4+}$, $Cs^+$ and $FA^+$. Table S4 summarizes the corresponding elemental ratios, including $I^-/(Pb^{2+} + Sn^{2+})$, $Sn^{2+}/Pb^{2+}$, $I^-/Sn^{2+}$, $I^-/Pb^{2+}$ and $Cs^+/FA^+$ ratios. Additionally, the $Sn^{4+}/Sn^{2+}$ ratio provides a measure of the relative amount of Sn oxidized on the surface.

For the bulk perovskite composition $FA_{0.75}Cs_{0.25}Pb_{0.5}Sn_{0.5}I_3$, the expected $I^-/(Pb^{2+} + Sn^{2+})$ and $Sn^{2+}/Pb^{2+}$ ratios are 3 and 1, respectively. The values obtained for the reference surface deviate substantially from these expected ratios. As shown in Table S4, the reference surface exhibits a highly nonstoichiometric surface with $I^-/(Pb^{2+} + Sn^{2+})$ and $Sn^{2+}/Pb^{2+}$ratios of 2.2 and 2.5, respectively. This testifies to a significant excess of Sn relative to Pb on the untreated surface. Following $EDAI_2$ treatment, the $I^-/(Pb^{2+} + Sn^{2+})$ ratio increases from 2.2 to 2.6, while the $Sn^{2+}/Pb^{2+}$ ratio decreases from 2.5 to 2.1. Although these changes indicate an improvement relative to the untreated surface, the surface composition remains substantially different from the bulk stoichiometry.

The most pronounced change is observed after EDA-treatment. Here, the $I^-/(Pb^{2+} + Sn^{2+})$ ratio increases to 3.2, closer to the expected value of 3, while the $Sn^{2+}/Pb^{2+}$ ratio decreases to 1.2, approaching the expected value of 1. EDA treatment therefore reconstructs the initially Sn-rich surface towards a composition substantially closer to the nominal perovskite stoichiometry than either the reference or $EDAI_2$-treated surfaces.

The $I^-/Sn^{2+}$ and $I^-/Pb^{2+}$ ratios provide further insight into the coordination of iodide relative to the two B-site cations. For the EDA-treated surface, these ratios are 5.8 and 7.1, respectively, indicating a more even coordination. In contrast, the $EDAI_2$-treated surface exhibits $I^-/Sn^{2+}$ and $I^-/Pb^{2+}$ ratios of 3.8 and 8.2, respectively, while the corresponding values for the reference surface are 3.0 and 7.5. The substantially different $I^-/Sn^{2+}$ and $I^-/Pb^{2+}$ ratios for the reference and $EDAI_2$-treated surfaces further reflect their Sn-rich compositions, whereas the EDA-treated surface exhibits ratios more consistent with the near-equimolar abundance of Sn and Pb.

The $Cs^+/FA^+$ ratio also changes upon treatment. The reference surface exhibits a $Cs^+/FA^+$ ratio of 0.7, which increases moderately to 0.9 following EDA treatment. In contrast, $EDAI_2$ treatment results in a substantially larger increase in the $Cs^+/FA^+$ ratio, from 0.7 to 1.8.

Finally, the degree of surface oxidation was assessed using the $Sn^{4+}/Sn^{2+}$ ratio. The ratio remains low for all three samples, with values of 0.05, 0.05 and 0.04 for the reference, $EDAI_2$-treated and EDA-treated sample surfaces, respectively. This indicates that there is only a minor contribution from oxidized $Sn^{4+}$ state.

**Supplementary Table 3** Atomic percentages for all relevant components of the various Sn-Pb perovskite films.

| Atomic % | EDA-treated | $EDAI_2$-treated | Reference |
|---|---|---|---|
| $I^-$ | 59.8 ± 1.1 | 51.6 ± 1.0 | 53.6 ± 1.0 |
| $Pb^{2+}$ | 8.4 ± 0.3 | 6.3 ± 0.2 | 7.1 ± 0.3 |
| $Sn^{2+}$ | 10.3 ± 0.5 | 13.5 ± 0.7 | 17.7 ± 0.9 |
| $Sn^{4+}$ | 0.4 ± 0.1 | 0.7 ± 0.2 | 0.9 ± 0.2 |
| $Cs^+$ | 3.3 ± 0.1 | 5.7 ± 0.2 | 4.0 ± 0.2 |
| $FA^+$ | 3.7 ± 0.6 | 3.2 ± 0.5 | 5.9 ± 0.9 |

**Supplementary Table 4** Elemental ratios derived from the atomic percentages of the various Sn-Pb perovskite films listed in Supplementary Table 3.

| Elemental ratios | EDA-treated | $EDAI_2$-treated | Reference |
|---|---|---|---|
| $I^-/(Sn^{2+}+Pb^{2+})$ | 3.2 ± 0.2 | 2.6 ± 0.2 | 2.2 ± 0.2 |
| $Sn^{2+}/Pb^{2+}$ | 1.2 ± 0.1 | 2.1 ± 0.1 | 2.5 ± 0.1 |
| $I^-/Sn^{2+}$ | 5.8 ± 0.3 | 3.8 ± 0.2 | 3.0 ± 0.2 |
| $I^-/Pb^{2+}$ | 7.1 ± 0.3 | 8.2 ± 0.3 | 7.5 ± 0.3 |
| $Sn^{4+}/Sn^{2+}$ | 0.04 ± 0.02 | 0.05 ± 0.02 | 0.05 ± 0.02 |
| $Cs^+/FA^+$ | 0.9 ± 0.1 | 1.8 ± 0.3 | 0.7 ± 0.1 |

**Supplementary Table 5** Probing depth for the different core levels.

| Core level | Kinetic energy (eV) | Inelastic mean free path (nm) | Information depth (nm) | #Unit cells probed |
|---|---|---|---|---|
| I 3*d* | 867 | 2.2 | 6.6 | 10.5 |
| Pb 4*f* | 1348 | 2.4 | 7.2 | 11.4 |
| Sn 3*d* | 1000 | 2.8 | 8.4 | 13.4 |
| Cs 3*d* | 762 | 2.1 | 6.3 | 10.0 |
| C 1*s* | 1202 | 2.6 | 7.8 | 12.4 |
| N 1*s* | 1087 | 2.5 | 7.5 | 11.9 |
| O 1*s* | 956 | 2.4 | 7.2 | 11.4 |

**Supplementary Text 3**

The information depth (95% signal contribution depth) was calculated based on the inelastic mean free path (IMFP) obtained from the TPP-2M model[10] using M=570 $gmol^{-1}$, ρ=4.5 $gcm^{-3}$, Eg=1.3 eV, and Nv=38. Based on the XRD data, the unit-cell height along the c-axis is c=0.629 nm. The probing depth of XPS follows the Beer-Lambert law: $I=I_0e^{-d/\lambda}$, where I is the detected photoelectron intensity, $I_0$ is the initial intensity, d is the depth below the surface, and λ is the inelastic mean free path (IMFP). As the signal intensity decays exponentially in depth, the majority of the detected XPS signal originates from the uppermost layers of the material.

For the Sn 3*d* core level, the information depth is estimated to be approximately 8.4 nm. Therefore, about 95% of the detected Sn 3*d* signal originates from the upper ~8.4 nm of the sample, corresponding to roughly 13-14 perovskite unit cells (based on a unit-cell height of 0.629 nm). This estimate assumes an ideal, contamination-free surface. In the presence of surface contamination or molecular overlayers, such as adventitious carbon or adsorbed species, the effective number of perovskite unit cells contributing to the detected signal will be reduced. The information depth varies slightly among the different core levels owing to differences in the kinetic energies of the emitted photoelectrons. However, these variations correspond to only a few unit cells and are systematic across all samples. Consequently, they do not significantly influence the compositional ratios discussed in the previous section. Small differences in the effective probing depth between samples may arise from variations in the thickness of the surface carbon overlayer and minor sample misalignment during measurement, with the latter estimated to be less than approximately 5°.

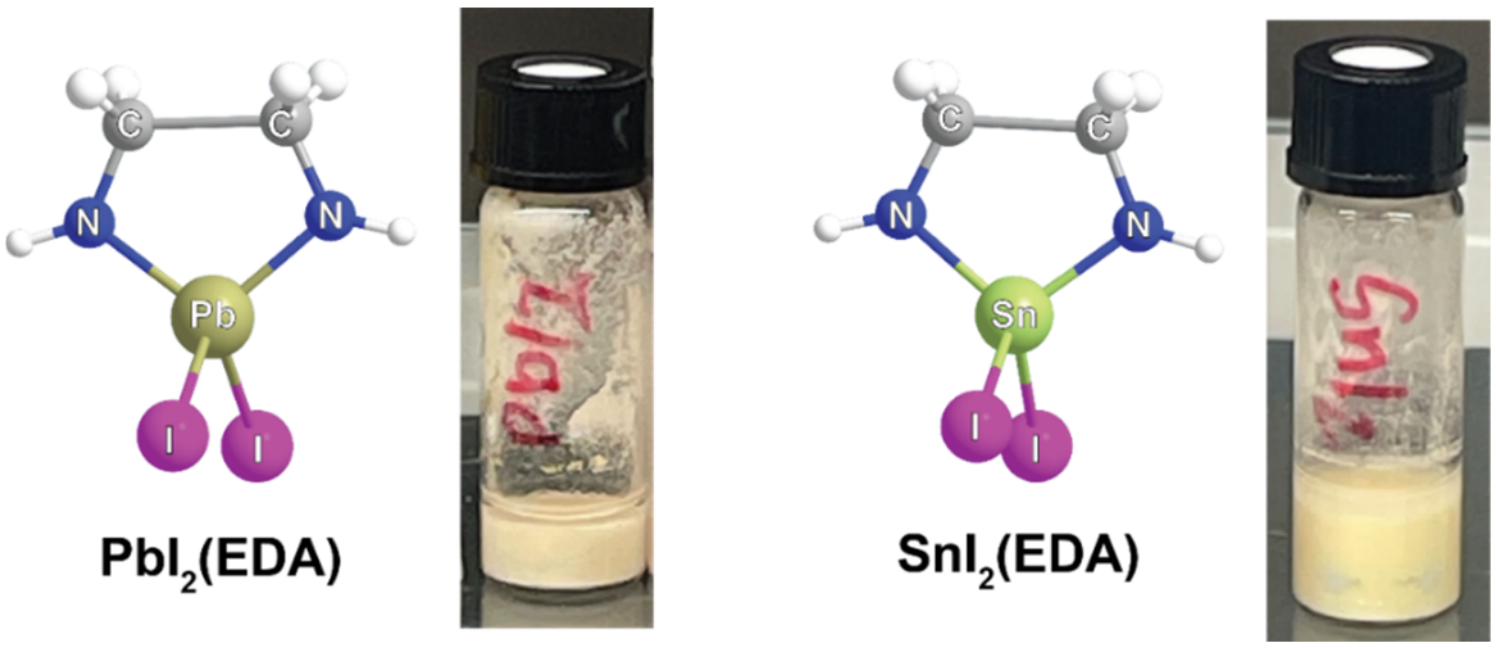

**Supplementary Fig. 15** $PbI_2$(EDA) and $SnI_2$(EDA) complexes in solution, ratio $MI_2$:EDA (1:1)

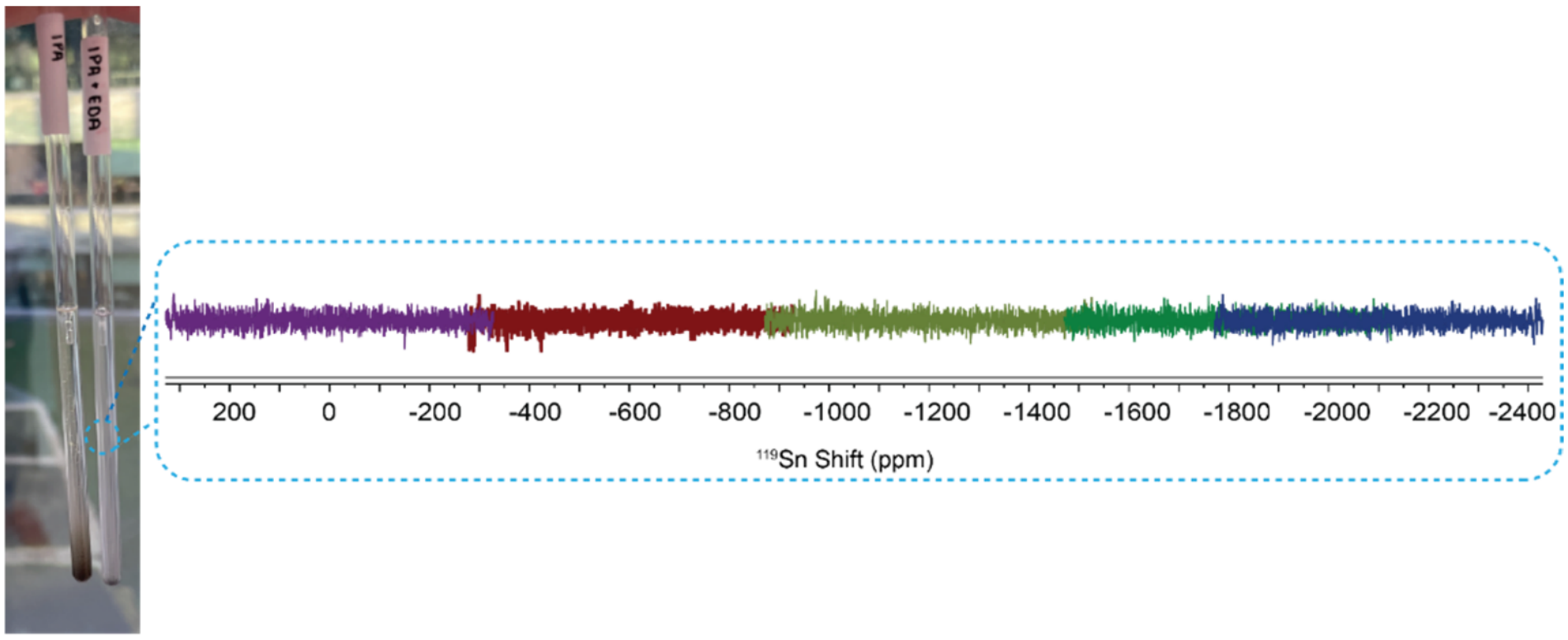

**Supplementary Fig. 16** $^{119}$Sn NMR of scratched Sn-Pb perovskite film in IPA-EDA solution

**Supplementary Text 4**

To examine the reaction between FAI and EDA proposed previously, and the possible formation of cyclic imidazolinium species in Pb-based perovskites,[11] we monitored FAI solutions containing EDA or $EDAI_2$ using Fourier-transform infrared spectroscopy (FTIR). Fig. S17 presents the spectra of FAI solutions mixed with EDA or $EDAI_2$. At the concentration used for device fabrication (0.1 mM), no discernible differences are observed between the EDA+FAI and $EDAI_2$+FAI solutions relative to pristine FAI, probably because of the low additive concentration and dominant solvent signals. In contrast, at a higher EDA concentration (2 mM), an additional feature emerges at ~1602 $cm^{-1}$ in the EDA+FAI solution, which can be attributed to N-H bending (scissoring/deformation) vibration of EDA primary amine groups, and is slightly shifted compared to pure EDA, suggesting an interaction between EDA and FAI. However, no C=N stretching band associated with imidazolinium formation is observed.

After post-treatment of our Sn-Pb perovskite films with high-concentration EDA (20 mM), XRD (Fig. S18) also reveals no additional diffraction peaks associated with imidazolinium, contrary to what has been observed for Pb perovskite,[11] suggesting a distinct response of Sn-Pb perovskites compared to their Pb-only counterparts. We further examined the interaction between the non-perovskite phase $Pb_{0.5}Sn_{0.5}I_2$ and the passivating molecules. Notably, the $EDAI_2$-treated film exhibits a pronounced color change, indicating chemical instability. In contrast, EDA treatment suppresses oxidation of the metal species and stabilizes $Pb_{0.5}Sn_{0.5}I_2$, as confirmed by FTIR spectra, UV-vis absorption spectra, and XRD patterns (Figs. S19 and S20).

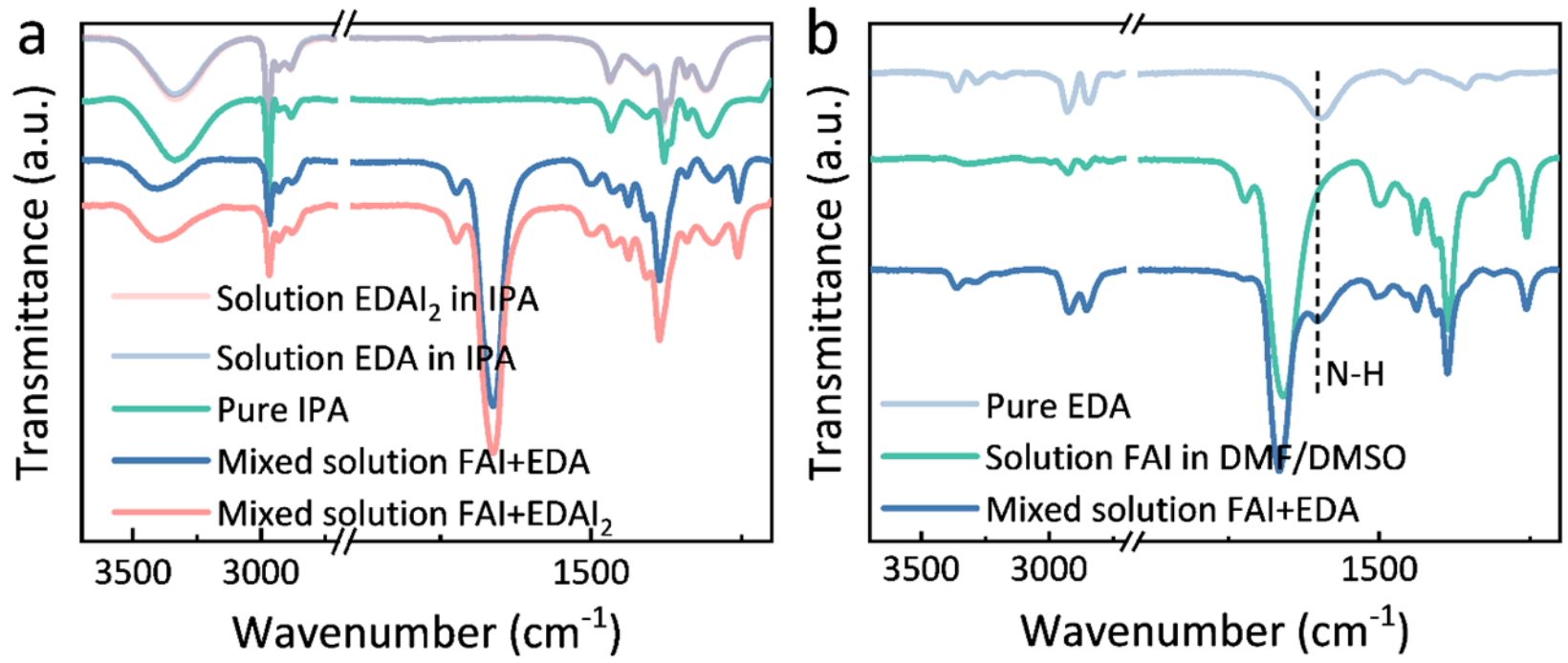

**Supplementary Fig. 17** (a) FTIR spectra of EDA, $EDAI_2$, and FAI mixed with EDA or $EDAI_2$ at the concentrations used for device fabrication. (b) FTIR spectra of pure EDA, FAI dissolved in DMF/DMSO, and FAI solutions containing 2 mM EDA, showing the interactions between EDA and the precursor solution.

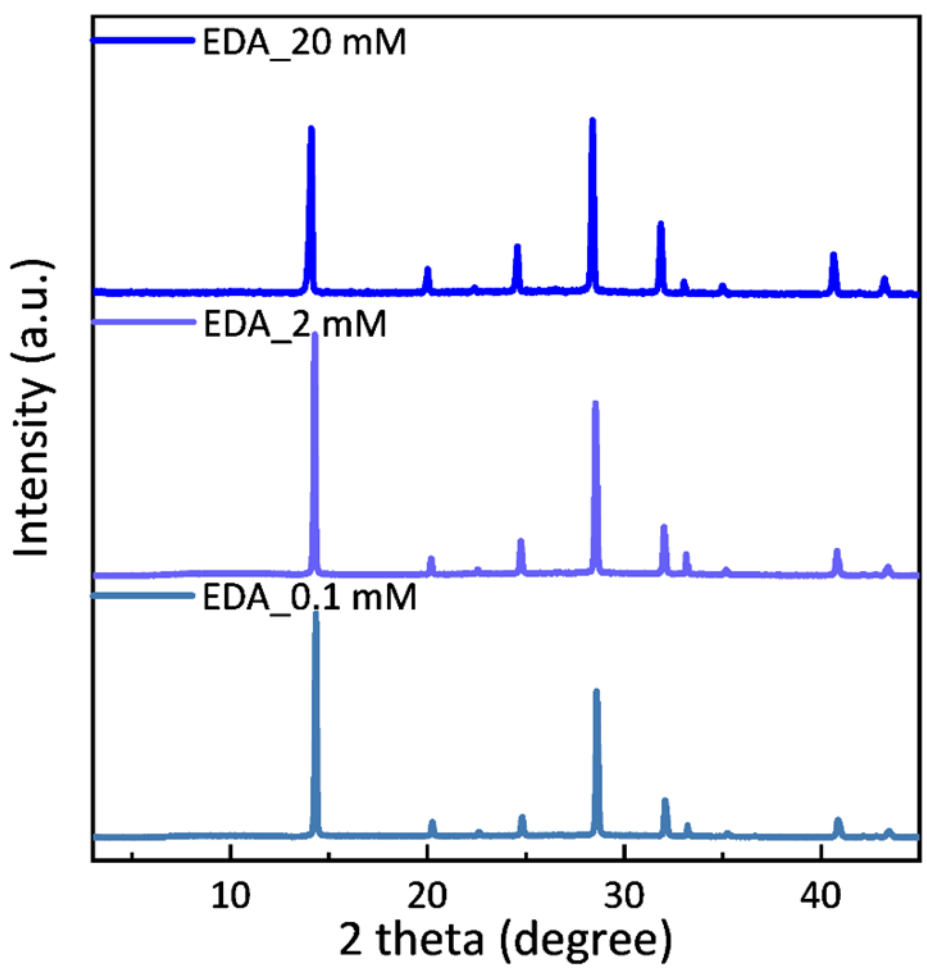

**Supplementary Fig. 18** XRD patterns of Sn-Pb perovskite films deposited on Si substrates and treated with three different EDA concentrations; no additional diffraction peaks are observed when the film was treated with high concentration EDA.

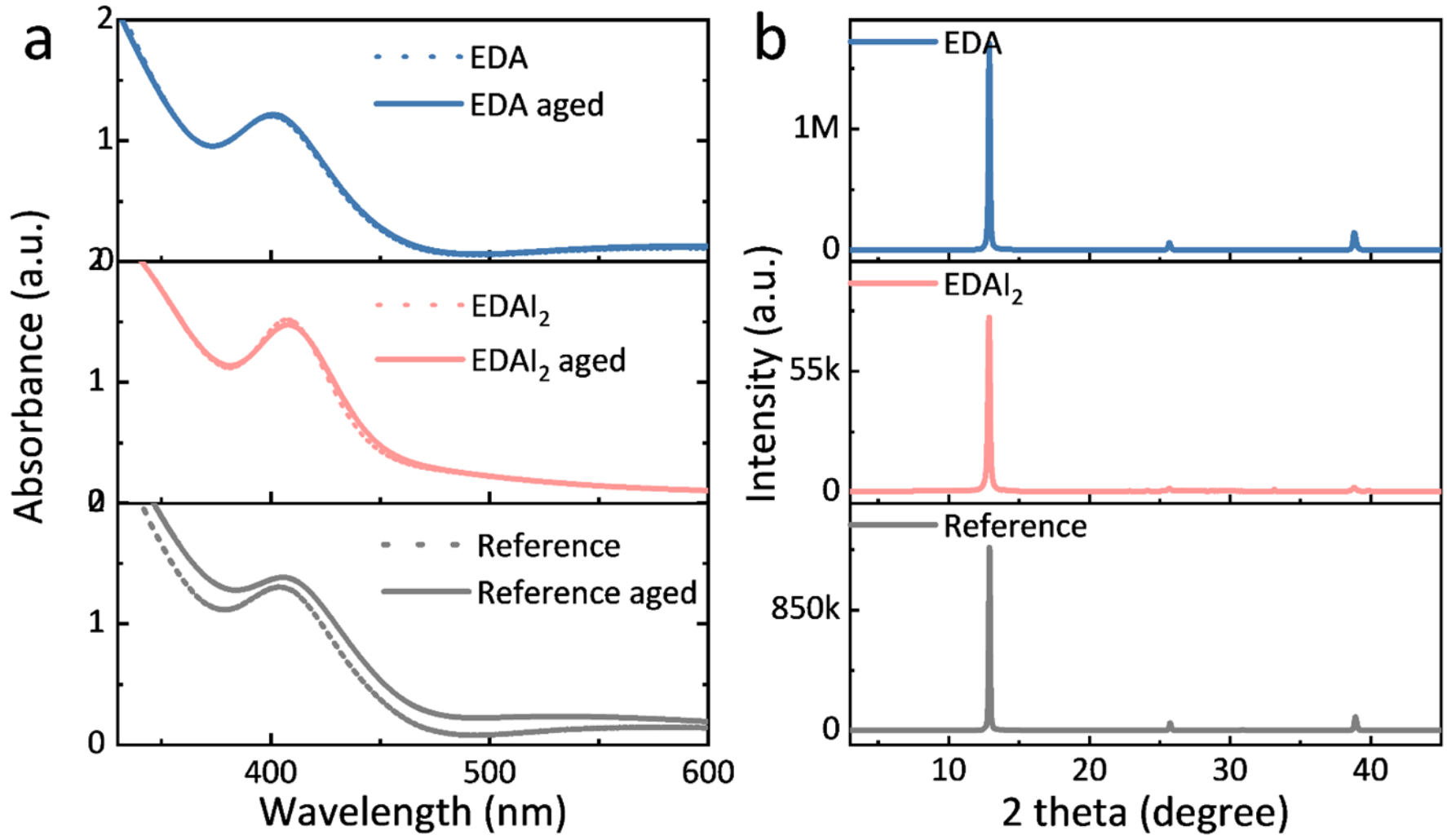

**Supplementary Fig. 19** (a) Time-dependent UV-vis absorption spectra of the reference, $EDAI_2$-passivated, and EDA-passivated Perovskite films. (b) The differences in diffraction peak intensity of $Sn_{0.5}Pb_{0.5}I_2$ films with EDA and $EDAI_2$ surface passivation.

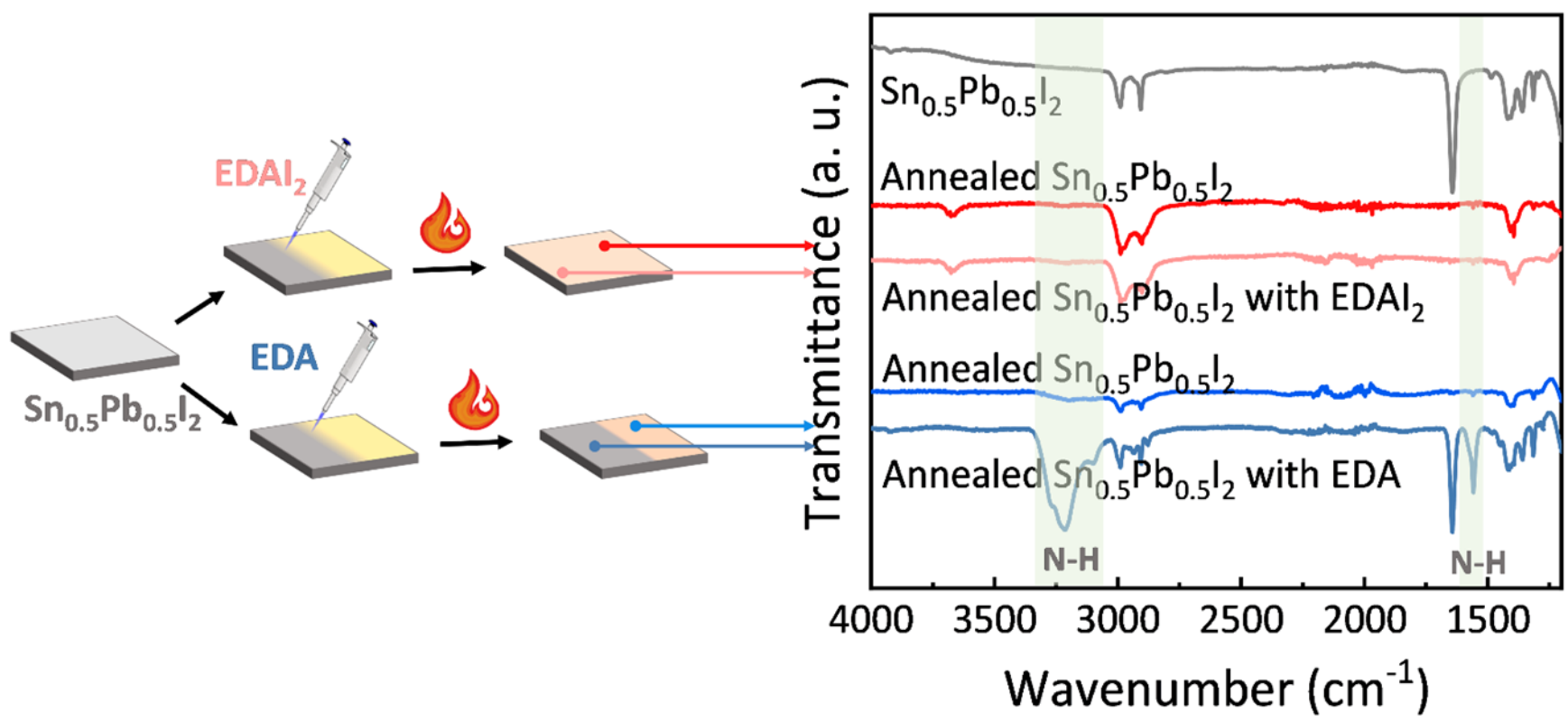

**Supplementary Fig. 20** FTIR spectra of $Sn_{0.5}Pb_{0.5}I_2$ films after EDA and $EDAI_2$ surface passivation. Following annealing, the untreated and $EDAI_2$-treated films exhibit pronounced color changes, indicative of severe chemical degradation and accelerated oxidation. In contrast, the EDA-treated film remains largely unchanged, underscoring the chelation-induced stabilization provided by EDA.

**Supplementary Text 5**

**Mixed Pb–Sn perovskite slab model**

Starting from the experimental bulk crystal structure of the Sn-Pb perovskite ($FA_{0.75}Cs_{0.25}Pb_{0.5}Sn_{0.5}I_3$), we prepared a symmetric 2x2 slab containing five metal-halide, see Fig. S21a.

Attention was devoted to the metal-terminated surface, which exposes undercoordinated Pb and Sn atoms and therefore represents the most relevant model for investigating the electronic defects associated with incomplete surface coordination. Starting from this pristine slab, we evaluated two different passivation strategies based on the adsorption of ethylenediamine, EDA, and its corresponding diammonium salt, $EDAI_2$. These two models were selected to directly compare the effects of neutral diamine and a conventional ammonium-based passivating agent on the structural and electronic properties of the mixed perovskite surface.

Following approaches reported previously,[12,13] electronic-structure analysis (Figs. S21b,c) shows that both passivation treatments effectively suppress the surface trap states associated with the undercoordinated Pb and Sn atoms (Fig. S21a), restoring a substantially cleaner electronic structure in the vicinity of the band edges. Despite this common electronic effect, the two treatments induce markedly different structural responses.

For the diammonium salt, the organic cation passivates the exposed metal sites while forming a relatively ordered surface overlayer that resembles the organic-inorganic arrangement of the perovskite lattice (Fig. S21d), thereby largely preserving the underlying metal–halide framework. By contrast, adsorption of the neutral diamine induces a pronounced reconstruction of the surface. The amino groups directly coordinate the exposed Pb and Sn atoms, reorganizing the local metal-halide environment and producing substantial structural distortions. This strong surface reconstruction is consistent with the experimentally observed structural changes following EDA treatment (Fig. S21e).

The calculations therefore indicate that both ammonium- and amine-based treatments eliminate electronically detrimental surface states, but through fundamentally different atomistic mechanisms. The diammonium salt forms a conventional passivating overlayer, whereas diamine coordinates directly with undercoordinated surface metal atoms and induces more extensive surface reconstruction. To quantify the thermodynamic driving force associated with the two surface treatments, we calculated the passivation energies of EDA and $EDAI_2$ using the following expression:

$E_{tot} = ([E_{pass\text{-}slab} – (E_{pristine} – n^*E_{mol})])/n_m$

where $E_{tot}$ is the total energy of the passivated perovskite surface, $E_{pristine}$ is the energy of the pristine metal-terminated slab, and $E_{mol}$ is the energy of the isolated EDA or $EDAI_2$ species multiplied by n (the number of molecules/salts present in the passivation layer). The passivation energy is normalized dividing by $n_m$, the number of the undercoordinated surface metal sites. For EDA, the passivation energy amounts to 1.51 eV when one molecule chelates a surface Sn atom, compared with 1.32 eV for coordination to Pb. EDA therefore preferentially chelates Sn over Pb by 0.19 eV. $EDAI_2$ exhibits a similar but weaker selectivity, with passivation energies of 1.44 and 1.39 eV per site for Sn and Pb, respectively. These results indicate that both the neutral diamine and the diammonium species preferentially interact with undercoordinated Sn atoms, although the metal selectivity is considerably stronger for EDA. Moving to the full passivated models, we found that EDA and $EDAI_2$ show comparable passivation energies (1.47 and 1.61 eV per site, respectively).

**Modeling of the Deprotonation Process**

To investigate the possible deprotonation of the diammonium species at the perovskite surface, a reduced reactive model was constructed starting from the previously optimized metal-terminated slab. In particular, the uppermost two metal-halide layers were extracted from the complete five-layer slab. The atomic coordinates of the metal and halide atoms belonging to the lower portion of the reduced model were kept fixed at their optimized slab positions, whereas the atoms of the upper surface layer and the adsorbed molecular species were allowed to fully relax. This approach preserves the structural constraint exerted by the underlying perovskite lattice while allowing the surface region directly involved in the chemical process to reorganize.

An $EDAI_2$ salt was subsequently placed on the exposed metal-terminated surface, and the deprotonation process leading to the formation of the corresponding neutral amine and HI was investigated. The reaction was initially modelled on the pristine surface, in the absence of structural defects, and subsequently repeated for different defective configurations to evaluate the influence of the local surface environment on proton transfer.

We considered defective models including a surface iodide vacancy ($V_{I+}$) and iodide Frenkel defect ($V_{I+}$ and $I_I$-), alternatively. The latter was generated by removing one iodide ion from its original lattice position and placing the corresponding iodine species at an interstitial site in a different region of the model. In this way, the overall composition and charge neutrality of the system were preserved while introducing a vacancy-interstitial defect pair.

The same deprotonation process was therefore evaluated on the pristine surface and on each defective model, allowing the effect of metal and halide defects on the thermodynamic feasibility of HI formation to be directly assessed. The energetics of the reaction in presence of the iodine vacancy and in presence of the Frenkel defect pair ($V_{I+}/I_{I-}$), see Fig. S21g, gives the same energetics reported in the main text.

**Computational Details**

All calculations were carried out using Quantum Espresso program package[14] along with PBE functional[15] and ultra soft pseudo potential with electrons from I 6*s*, 6*p*; Pb 6*s*, 6*p*, 5*d*; Sn 5*s*, 5*p*, 4*d*; C, N, 2*s*, 2*p*; Cs 5*s*, 5*p*, 6*s*; H 1*s* shells explicitly included. D3 dispersion interaction was considered. Plane-wave basis set cutoffs for the smooth part of the wave functions and for the density of 25 and 200 Ry, respectively, were chosen. Supercell calculations were performed at Γ point. Experimental cell parameters were adopted. Car-Parrinello molecular dynamic simulations were carried out using the same setup above, with simulation time step of dt=10 a.u. and employing all atomic masses equal to 5.0 a.m.u. along with electron mass of 1000. The structure models adopted for the molecular dynamic simulation were the same used for the deprotonation processes described above in a full coverage condition of EDA and $EDAI_2$, alternatively. A proper initial ions position randomization from the optimized geometries was adopted to achieve ~350K.

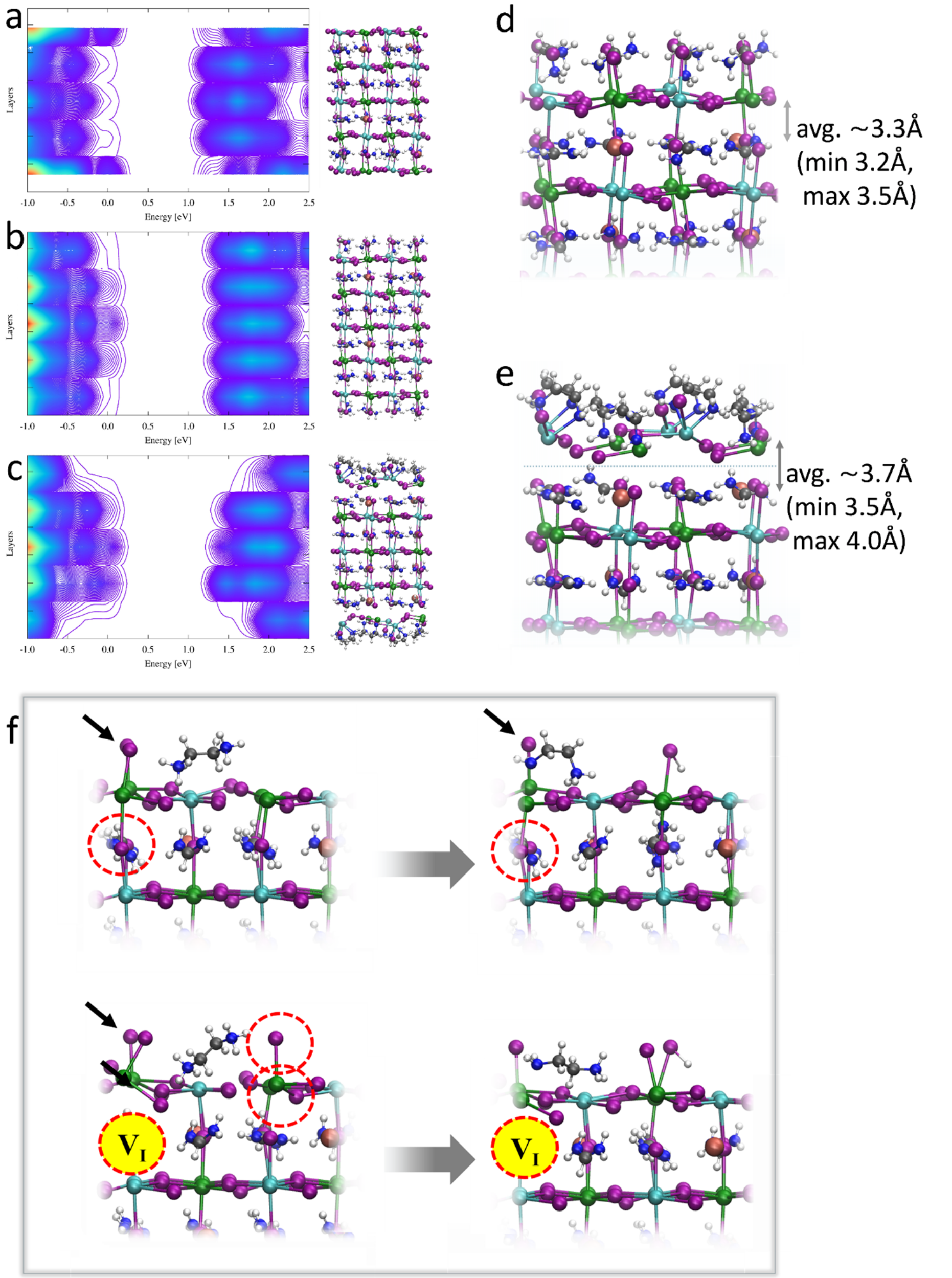

**Supplementary Fig. 21** Isodensity contour plots of DOS layer by layer for the (a) bare metal-terminated slab, the (b) relative $EDAI_2$ passivated and the (c) EDA passivated surface. The detailed of the geometry structure for the (d) $EDAI_2$- and (e) EDA-passivated slab. (f) The deprotonation process of $EDAI_2$ leading to the formation of HI in presence of the Frenkel defect pair ($V_{I+}/I_{I-}$); the black arrows indicate the $I_{I-}$ species.

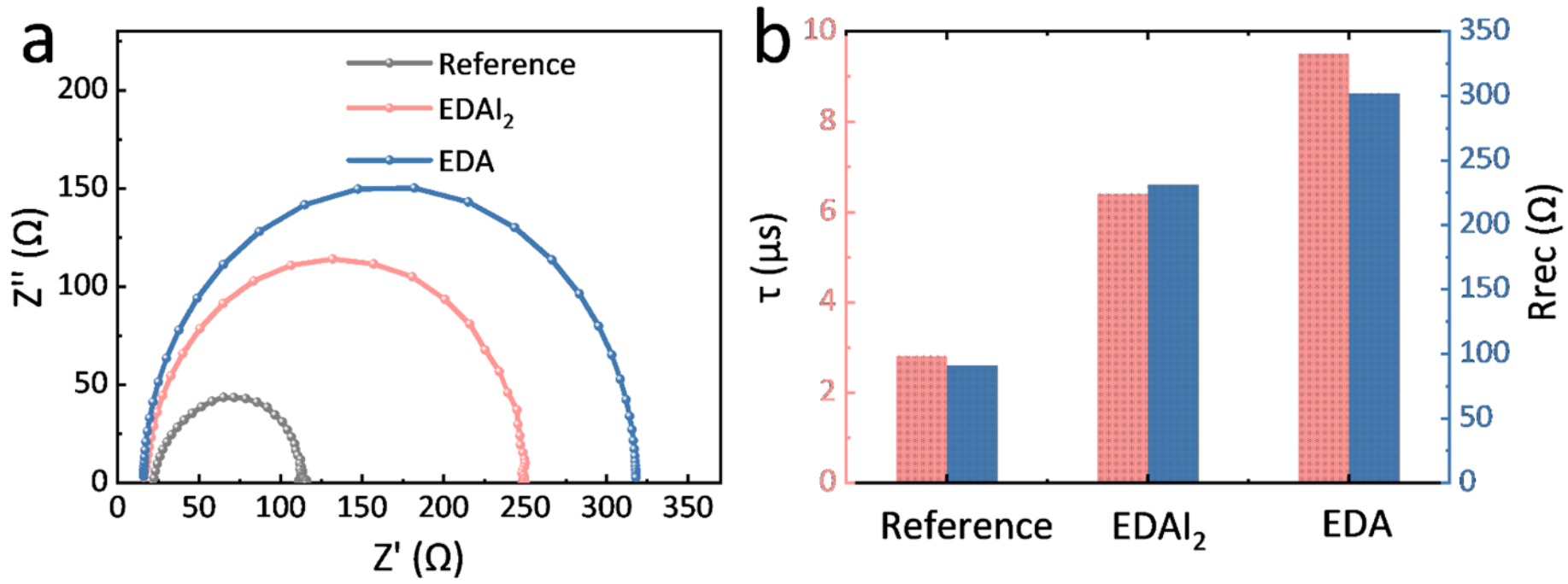

**Supplementary Fig. 22** (a) Nyquist plots of devices prepared with reference, $EDAI_2$-, and EDA-treated Sn-Pb perovskite, measured under open-circuit voltage conditions. (b) Recombination resistance and carrier lifetime of corresponding devices extracted from the Nyquist plots.

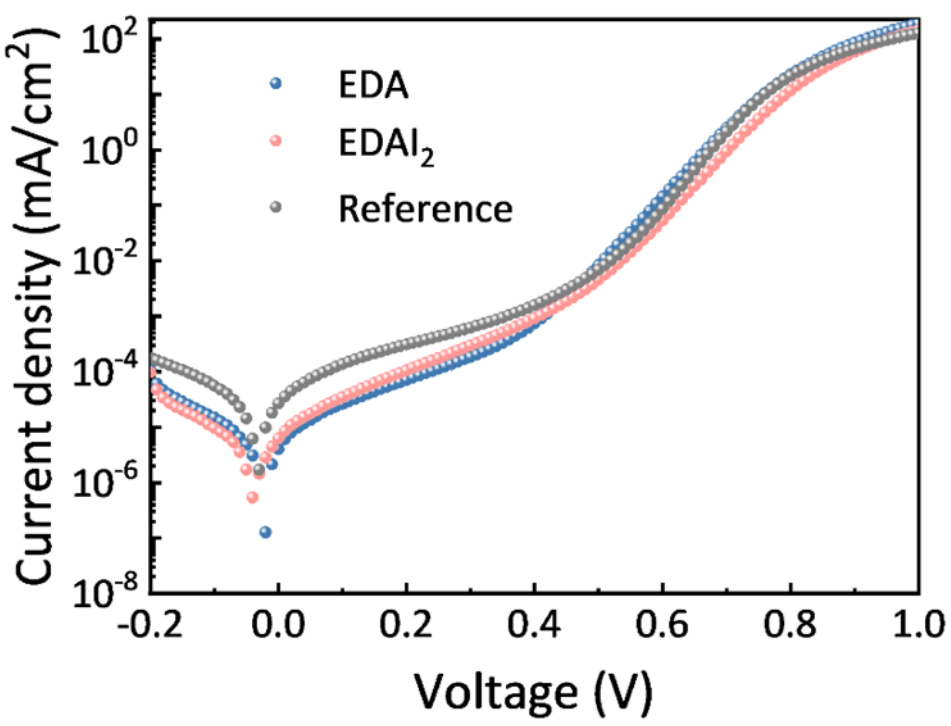

**Supplementary Fig. 23** Dark J-V characteristics of devices fabricated with the reference film and with films subjected to EDA and $EDAI_2$ passivation treatments.

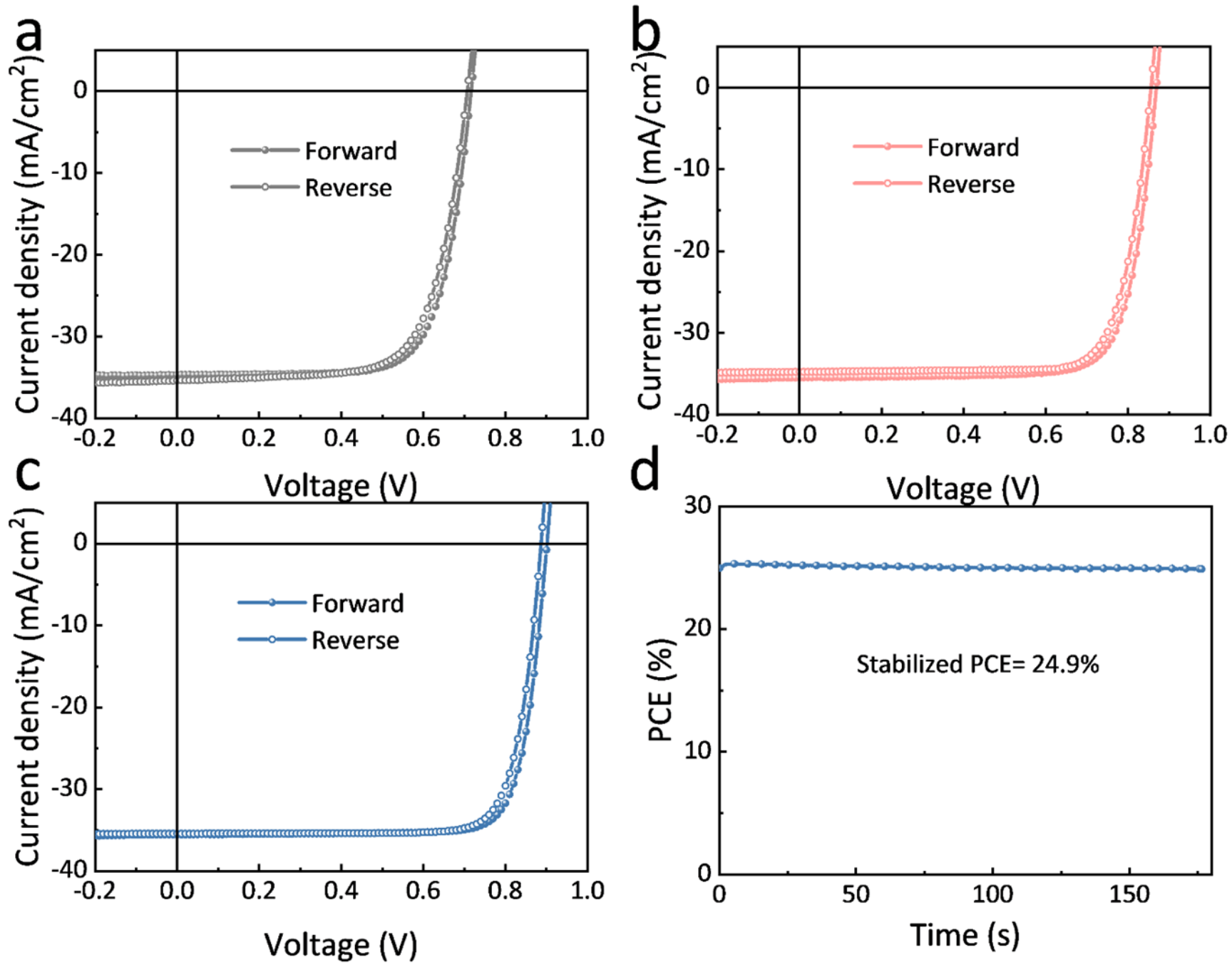

**Supplementary Fig. 24** Forward and reverse J-V scans of the champion devices measured under AM 1.5G illumination: (a) reference, (b) $EDAI_2$-treated, and (c) EDA-treated Sn-Pb perovskites. (d) Stabilized power output of the EDA-treated champion device.

**Supplementary Table 6** Summary of reported Sn-Pb photovoltaic devices with PCEs exceeding 24% in literature.

| Perovskite | Surface passivation | PCE (%) | References |
|---|---|---|---|
| $Cs_{0.2}FA_{0.8}Sn_{0.5}Pb_{0.5}I_3$ | yes | 24.07 | [16] |
| $FA_{0.7}MA_{0.3}Pb_{0.5}Sn_{0.5}I_3$ | no | 24.23 | [17] |
| $Rb_{0.04}Cs_{0.2}FA_{0.76}Pb_{0.5}Sn_{0.5}I_3$ | yes | 24.5 | [18] |
| $Cs_{0.05}FA_{0.7}MA_{0.25}Pb_{0.5}Sn_{0.5}I_3$ | yes | 24 | [19] |
| $FA_{0.7}MA_{0.3}Pb_{0.5}Sn_{0.5}I_3$ | yes | 24.9 | [20] |
| $FA_{0.7}MA_{0.3}Pb_{0.5}Sn_{0.5}I_3$ | yes | 24.17 | [21] |
| $(FASnI_3)_{0.6}(MAPbI_3)_{0.4}$ | yes | 24.02 | [22] |
| $Cs_{0.1}FA_{0.6}MA_{0.3}Sn_{0.5}Pb_{0.5}I_3$ | no | 24.3 | [23] |
| $FA_{0.65}MA_{0.3}Cs_{0.05}Pb_{0.5}Sn_{0.5}I_3$ | yes | 24.2 | [24] |
| $FA_{0.6}MA_{0.3}Cs_{0.1}Pb_{0.5}Sn_{0.5}I_3$ | yes | 24.03 | [25] |
| $FA_{0.7}MA_{0.2}Cs_{0.1}Pb_{0.5}Sn_{0.5}I_3$ | no | 24.05 | [26] |
| $Rb_{0.04}Cs_{0.2}FA_{0.76}Pb_{0.5}Sn_{0.5}I_3$ | yes | 24.20 | [27] |
| $(FASnI_3)_{0.6}(MAPbI_3)_{0.4}$ | yes | 24.06 | [28] |
| $Cs_{0.05}MA_{0.35}FA_{0.5}Pb_{0.5}Sn_{0.5}I_3$ | yes | 24.07 | [29] |
| $FA_{0.7}MA_{0.3}Pb_{0.5}Sn_{0.5}I_3$ | yes | 24.19 | [30] |
| $FA_{0.7}MA_{0.3}Pb_{0.5}Sn_{0.5}I_3$ | yes | 24.11 | [31] |
| $Cs_{0.1}FA_{0.6}MA_{0.3}Pb_{0.5}Sn_{0.5}I_3$ | no | 24.10 | [32] |
| $Cs_{0.1}FA_{0.6}MA_{0.3}Pb_{0.5}Sn_{0.5}I_3$ | yes | 24.08 | [33] |
| $FA_{0.7}MA_{0.3}Pb_{0.5}Sn_{0.5}I_3$ | yes | 24.01 | [34] |
| $FA_{0.75}Cs_{0.25}Pb_{0.5}Sn_{0.5}I_3$ | yes | 25.86 | Our work |

**Supplementary Table 7** Photovoltaic parameters of the champion devices prepared with the reference, the EDA-treated and the $EDAI_2$-treated Sn-Pb perovskite films.

| Type of device | Scan direction | $V_{OC}$ (V) | $J_{SC}$ (mA/cm$^2$) | FF (%) | PCE (%) |
|---|---|---|---|---|---|
| EDA-treated | Forward | 0.901 | 35.56 | 81 | 25.86 |
| | Reverse | 0.887 | 35.42 | 80 | 25.19 |
| $EDAI_2$-treated | Forward | 0.869 | 35.46 | 77 | 23.86 |
| | Reverse | 0.856 | 34.74 | 78 | 23.17 |
| Reference | Forward | 0.716 | 34.71 | 73 | 18.11 |
| | Reverse | 0.707 | 35.37 | 70 | 17.44 |

**Supplementary Table 8** Mean values and standard deviations of photovoltaic parameters (PCE, FF, Voc, and Jsc) for perovskite solar cells.

| Type of device | $V_{OC}$ (V) | $J_{SC}$ (mA/cm$^2$) | FF (%) | PCE (%) |
|---|---|---|---|---|
| EDA-treated | 0.887 ± 0.012 | 35.04 ± 0.53 | 79.39 ± 0.80 | 24.69 ± 0.70 |
| $EDAI_2$-treated | 0.851 ± 0.008 | 34.81 ± 0.55 | 76.05 ± 1.06 | 22.55 ± 0.63 |
| Reference | 0.661 ± 0.034 | 33.85 ± 0.96 | 70.15 ± 1.14 | 15.68 ± 0.96 |

**Supplementary Text 6**
**Spectral mismatch factor**

Fig. S25 shows the irradiance spectrum of the simulator used for the EQE measurements. A spectral mismatch factor (M) was applied to correct for differences between the spectrum of the white light used during EQE measurements and the AM1.5G reference spectrum. The standard energy-responsivity formulation (equivalent to the photon-EQE form) was used, where, given the spectral irradiance $E(\lambda)$ and the spectral responsivity $SR(\lambda)$, the mismatch factor is

$$M=\frac{\int E_{AM1.5}(\lambda)SR_{RefCell}(\lambda)d\lambda}{\int E_{AM1.5}(\lambda)SR_{TestCell}(\lambda)d\lambda}\times\frac{\int E_{Simulator}(\lambda)SR_{TestCell}(\lambda)d\lambda}{\int E_{Simulator}(\lambda)SR_{RefCell}(\lambda)d\lambda}$$

Where $E_{AM1.5}(\lambda)$ is the irradiance of AM1.5G reference spectrum, and $E_{Simulator}(\lambda)$ is the irradiance of the actual lamp spectrum used during the EQE measurement, $SR_{RefCell}(\lambda)$ and $SR_{RefCell}(\lambda)$ denote the spectral responsivities of the reference Si cell and the test perovskite solar cell, respectively. The calculated mismatch factor M for the reported cells is 1.07, indicating that approximately 7% of the observed Jsc discrepancy originates from spectral mismatch.

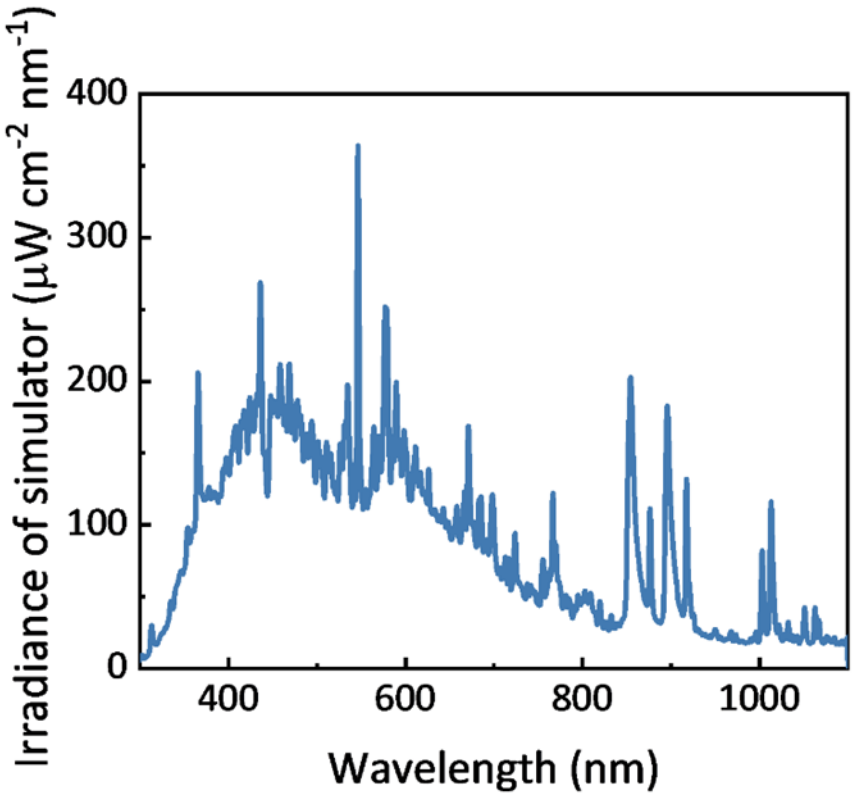

**Supplementary Fig. 25** Irradiance spectrum of the simulator used for the EQE measurements.

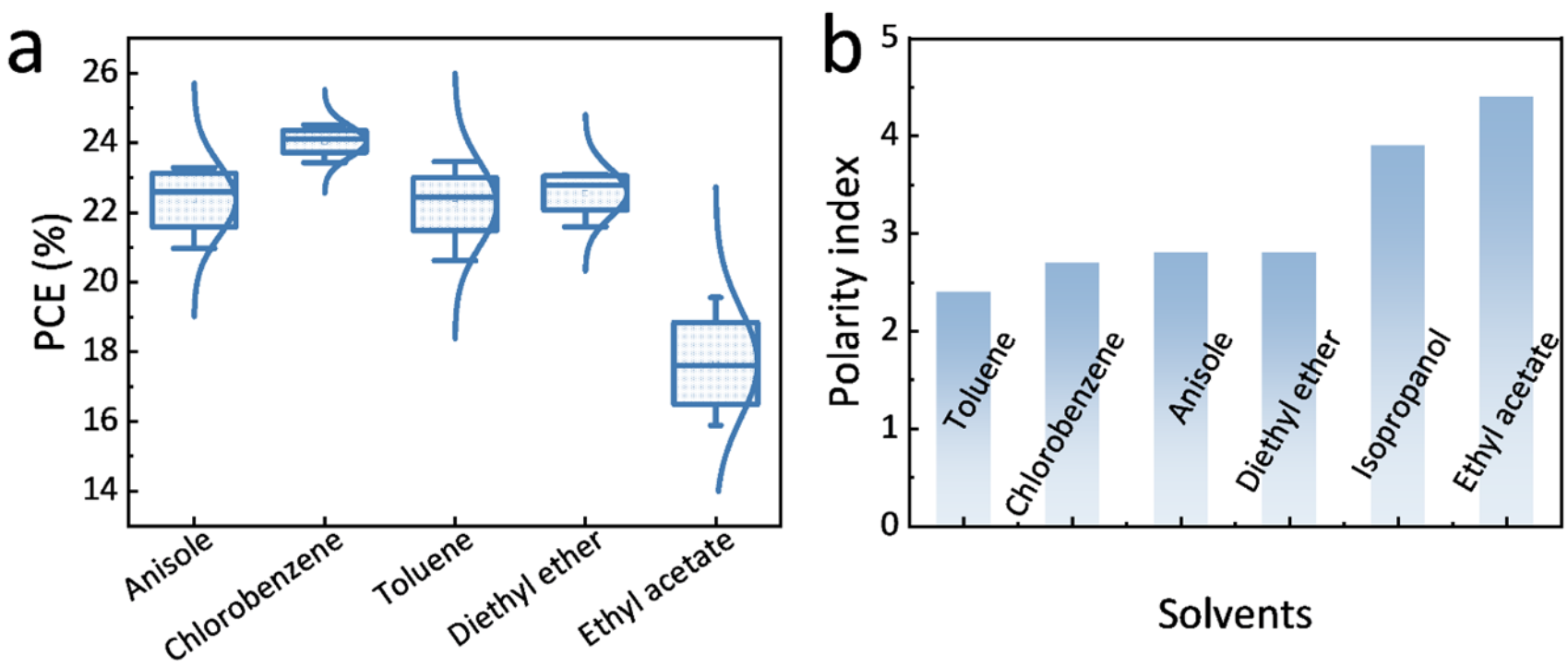

**Supplementary Fig. 26** (a) Comparison of the PCE of devices treated with EDA using different solvents. (b) Polarity index of the solvents used for EDA treatment.

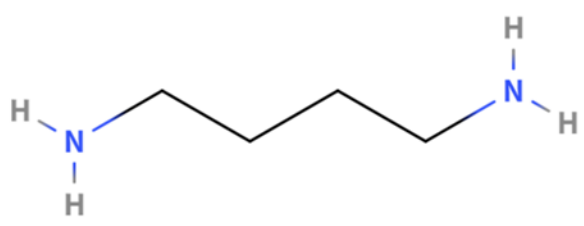

1,4-butanediamine (BDA)

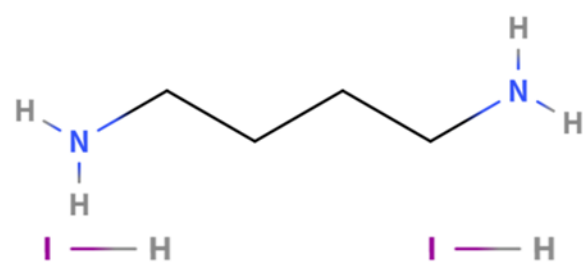

1,4-butanediamine dihydriodide ($BDAI_2$)

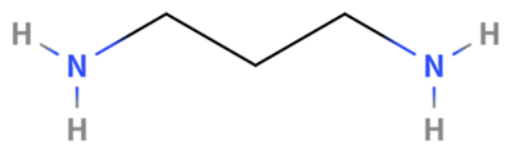

1,3-Propanediamine (PrDA)

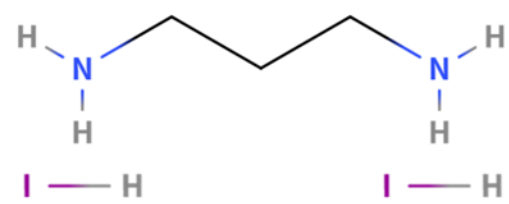

1,3-Propanediamine dihydriodide ($PrDAI_2$)

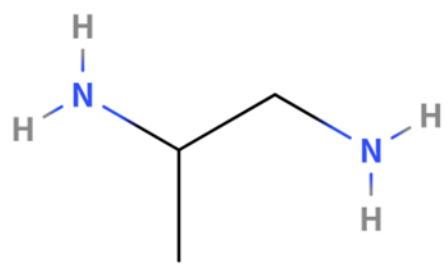

1,2-diaminopropane (DAP)

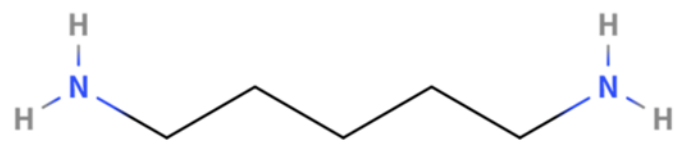

1,5-pentanediamine (PeDA)

**Supplementary Fig. 27** Molecular structures of a series of diamines and diammonium salts with different chain lengths.

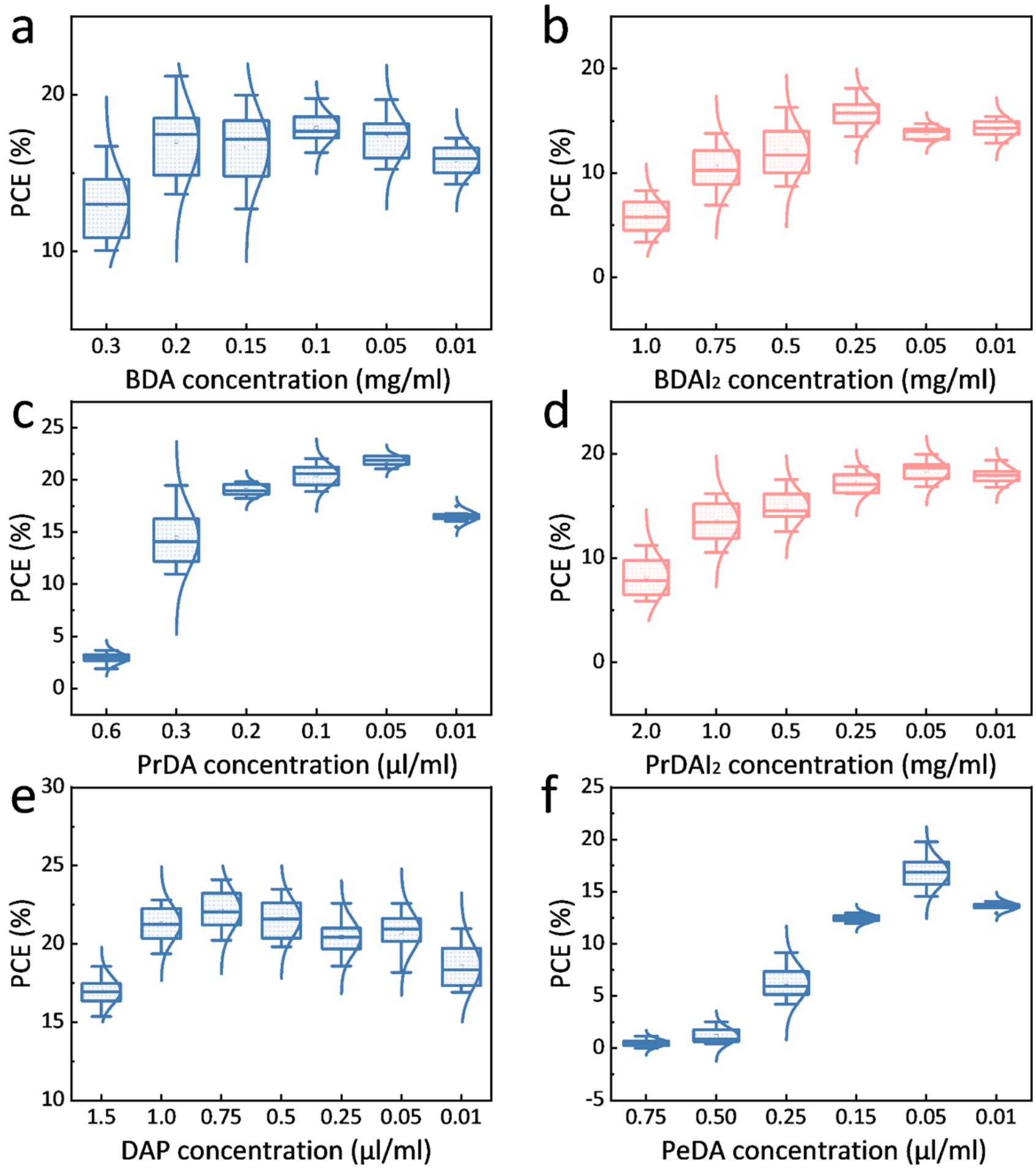

**Supplementary Fig. 28** Variation of PCE for solar cells fabricated with Sn-Pb perovskite films treated with different concentrations of diamines and with diammonium salts of different chain lengths. (a) 1,4-butanediamine, (b) 1,4-butanediamine dihydriodide, (c) 1,3-Propanediamine, (d) 1,3-Propanediamine dihydriodide, (e) 1,2-diaminopropane, and (f) 1,5-pentanediamine.

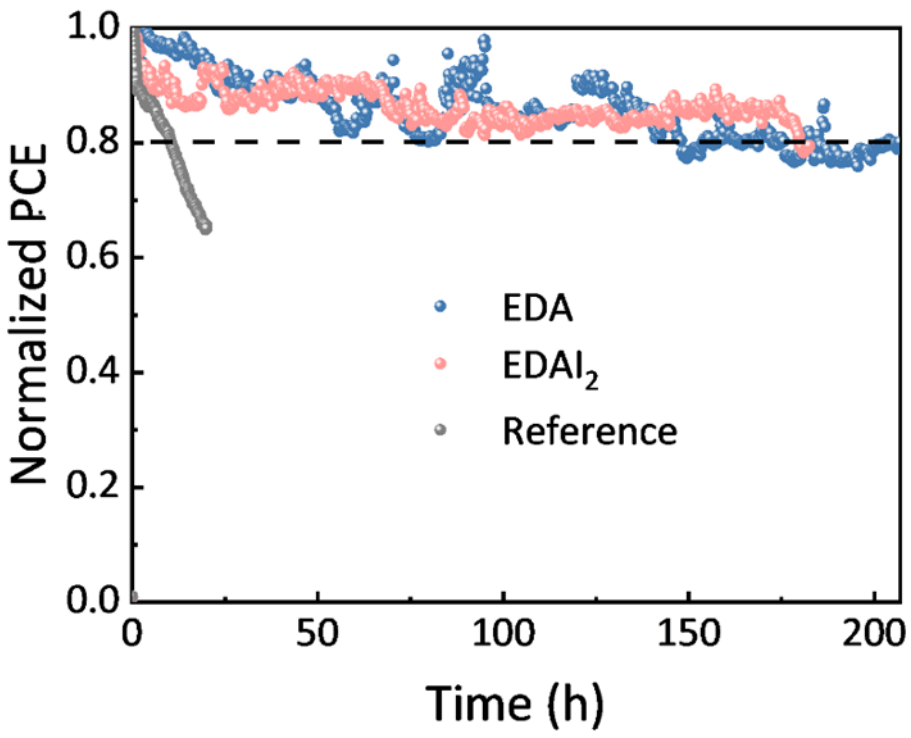

**Supplementary Fig. 29** Maximum power point tracking of devices fabricated with reference, EDA-treated, and EDAI$_2$-treated films under AM 1.5G illumination at 15-30 °C.